\documentclass[%
reprint,
nofootinbib,
amsmath,amssymb,
aps
]{revtex4-2}

\usepackage{subcaption}

\usepackage{graphicx}
\usepackage{dcolumn}
\usepackage{bm}
\usepackage{amsmath}
\usepackage{mathtools}

\usepackage{graphicx}
\usepackage{dcolumn}
\usepackage{bm}
\usepackage{amsmath}
\usepackage{mathtools}

\begin{document}
	
	\title {On the thermodynamic geometry of one-dimensional spin-3/2 lattice models}

	\author{Riekshika Sanwari}
	\email{riekshikas.phd19.ph@nitp.ac.in }
	\affiliation{National Institute of Technology, Patna}

	\author{Soumen Khatua}
	\email{soumenk.phd20.ph@nitp.ac.in }
	\affiliation{National Institute of Technology, Patna}

	\author{Anurag Sahay}
	\email{anuragsahay@nitp.ac.in}
	\affiliation{National Institute of Technology, Patna}

\begin{abstract}

Four-dimensional state space geometry is worked out for the exactly solved one-dimensional spin-3/2 lattice with a Blume-Emery-Griffiths (BEG) Hamiltonian as well as a more general one with a term containing a non-zero field coupling to the octopole moments. The phase behaviour of the spin-3/2 chain is also explored extensively and novel phenomena suggesting anomalies in the hyperscaling relation and in the decay of fluctuations are reported for a range of parameter values. Using the method of constrained fluctuations  worked out earlier in \cite{asknbads,riekan1} three sectional curvatures and a $3d$ curvature are obtained and shown to separately encode dipolar, quadrupolar and octopolar correlations both near and away from pseudo-criticality. In all instances of a seeming hyperscaling violation the $3d$ scalar curvature is found to encode the correlation length while the relevant $2d$ curvature equals the inverse of singular free energy. For parameter values where the order parameter fluctuation anomalously decays despite a divergence in correlation length the relevant scalar curvature undergoes a sign change to positive values, signalling a possible change in statistics.

\end{abstract}

	\maketitle

\section{Introduction}
\label{introthrhalf}

	Thermodynamic geometry (TG) has been extensively used to investigate phase transition phenomena in a wide class of physical systems ranging from fluids to black holes \cite{ruppajp}. By  introducing a Riemannian distance measure in the equilibrium state space TG forges a remarkable connection between the geometric invariants of the state space manifold and the underlying microscopic description of the physical system, thus providing a seemingly surprising thermodynamic route to statistical mechanical results. 
For example, upto a constant of order unity, the state space scalar curvature $R$ has been conjectured to be equal to the correlation volume in the vicinity of critical point \cite{rupp1, rupporiginal,ruppfluc} and, 
furthermore, it can be used to infer the coexistence curves in discontinous phase transitions as well as the Widom line in the supercritical phase \cite{maymaus, ruppantsr, deysark, ruppmay}. In addition to its well known correspondence with the correlation length both  near and away from criticality, the state space scalar curvature $R$ also possibly encodes some information of higher order statistics. 
 For instance, it has been widely observed that the sign of $R$ discriminates between repulsive and attractive nature of statistical interactions or between solid-like and fluid-like states of aggregation \cite{ruppmay,rupptrip, rupprep, behrouz1, behrouz2, ruppmac,paolo}.

There has been a sustained interest in the TG of exactly solved models, especially the one-dimensional lattice spin models. In spite of the absence of a finite temperature critical point the exactly solved models often exhibit sufficiently rich ground state phase structures which contain clues to understanding their higher dimensional counterparts. With analytical control at hand, they also provide fertile testing grounds for the predictions and claims of thermodynamic geometry. For example, the scalar curvature $R$ can be directly compared to the correlation length obtained via the transfer matrix of one-dimensional spin models. One of the first such models used to successfully verify TG is the one-dimensional Ising model which has a pseudocritical point at zero temperature in zero magnetic field. The scalar curvature earlier worked out numerically in \cite{ruppmag} was later found to be a surprisingly simple expression, \cite{mrugala}. Some other cases where TG has been applied to exactly solved models are the Ising model on a Bethe lattice, \cite{dolan1}, Ising model on planar random graphs, \cite{dolan2}, the spherical model, \cite{dolan3} and the one dimensional Potts model, \cite{dolan4} , decorated two-parameter Ising spin chain with frustration, \cite{ruppbelucci} and a ferromagnetic Ising model in an external magnetic field \cite{erdemr}. The TG of the one-dimensional spin-one model and of its mean field approximation was investigated by us earlier in \cite{riekan1} and \cite{riekan2}. See also \cite{erdemal,erdemal2} for an alternative approach to the geometry of the spin-one model.

With their rich phase structure comprising critical lines, coexisting surfaces and triple points,  higher spin lattice models have been of continued interest in the study of phase transition phenomena in systems with competing order parameters.   
One of the most popular spin-one models, the Blume-Emery-Griffiths (BEG) model, originally formulated to study the phase behaviour of $\mbox{He}^3$-$\mbox{He}^4$ mixtures, has  been used 
to study the phase behaviour in variety of systems, \cite{sivar,rys,microem,mukamel}. 
The spin-3/2 model, the system of interest in this work, was first used by Krinsky
and Mukamel \cite{krinm}  as an improvement  over the spin-1 lattice gas model of ternary fluid mixtures
of Mukamel and Blume \cite{mukamel}.  It was argued in \cite{krinm} that a true representation of the non-symmetric tricritical point found in ternary mixtures
requires a four dimensional parameter space which necessitates at least a spin-3/2 lattice. In addition to the dipolar and quadrupolar order parameters found in the spin-1 model the spin-3/2 case includes an additional octopolar order parameter. Thus, the Krinsky-Mukamel model is able to accommodate three different types of particles and a vacancy as required for a ternary mixture instead of just two
particles and a vacancy in the Mukamel-Blume model. Indeed, the mean field phase structure of the spin-3/2 lattice gas model gives a qualitatively correct picture of  multicritical phase
behaviour in ternary mixtures including the non-symmetric tricritical points found experimentally. On the other hand, a spin-3/2 BEG model containing only dipolar and quadrupolar interactions without crystal field was first used to model the phase behaviour of $DyVO_4$  \cite{sivad}. Other works include a study of the spin-3/2 BEG model with a crystal-field within the mean field approximation and also via a Monte Carlo simulation  \cite{barreto} and by taking the renormalization-group approach \cite{bakchich}. Antiferromagnetic spin-3/2 Blume Capel model was studied in \cite{bekhechi}.

Higher spin lattice models are especially interesting from a TG perspective. 
With their state space manifold of dimension three or more (eg, three dimensions in spin-1 and four in spin-3/2) they provide enough opportunity to explore higher dimensional thermodynamic geometry \cite{rupp8}. As against the two dimensional case where the only independent curvature is the state space scalar curvature $R$, now the full Riemannian curvature tensor can be exploited for more detailed information about underlying statistical interactions. In particular, curvatures on appropriately chosen slices of the higher dimensional thermodynamic manifold could be harnessed for information about specific order parameter correlations. In a basic form this approach of investigating curvatures on hypersurfaces was used in \cite{ruppknb,ruppknbfluc} in the context of Kerr Newman black holes and  in \cite{sssknbads} for Kerr-Newman AdS black holes. A formal mathematical basis termed the method of constrained fluctuations was worked out in \cite{asknbads} in the context of extended phase space Kerr-AdS black holes. The method was further developed in our earlier work \cite{riekan1} where it was applied to an ordinary thermodynamic system, namely the one-dimensional spin-one model. We were able to show in detail how  curvatures on suitable hypersurfaces separately encode correlations in the dipole and quadrupole fluctuations.

In this work we obtain the state space Riemannian geometry for the exactly solved one-dimensional spin-3/2 lattice model and extensively investigate its phase behaviour in the light of geometry. Three sectional curvatures and a $3d$ curvature are obtained by taking suitable hypersurfaces in the four dimensional state space manifold and their properties investigated vis-\'a-vis the fluctuations and correlations in order parameters. There are three order parameters in the model which gives rise to the possibility of two or more correlation lengths. We present our results on the extent to which geometry encodes the rich phase behaviour of the system. Further, we explore the connection between the sign changes in the curvatures and the change in pattern of fluctuations in order parameters. 

This paper is organised as follows. In section \ref{threeh} we obtain the most general Hamiltonian of a spin-3/2 lattice with nearest -neighbour interactions and then discuss its BEG limit and a more general limit where the octopolar field is non-zero. Restricting to the spin-3/2 chain we obtain the free energy, fluctuations moments and correlation lengths via its transfer matrix. In section \ref{phastru} we discuss in detail the ground state phase behaviour of the  spin-3/2 chain both for the BEG case and the more general case. Several new results are reported.
In section \ref{threehgeo} we briefly outline our development of the relevant spin-3/2 curvatures via the method of constrained fluctuations mentioned earlier. In section \ref{free} we report the asymptotic expressions of the singular free energy and the correlation lengths in different parameter regimes. In section \ref{geometry} we present the results of our detailed investigations into the state space geometry of the spin-3/2 chain.
Finally, in the concluding section \ref{conclu32} we summarize our key results and try to define the scope of our work.

\section{One-dimensional spin-3/2 model }
\label{threeh}

 Following \cite{krinm} and adding some explanation of our own, we first briefly outline a justification for the most general form of the spin-3/2 Hamiltonian from a ternary mixture lattice gas perspective. Let $S=3/2,1/2,-1/2$ represent, respectively, particle $1$, $2$, and $3$ and $S=-3/2$ represent a vacancy or particle $0$. The interaction between the particles is given by the coupling strengths $K_{11},K_{22},K_{33},K_{12},K_{13},K_{23}$ where the subscripts indicate the type of particles. The Hamiltonian can then be written in terms of projection operators (functions),
\begin{equation}
	\mathcal{H}=-\sum_{\langle ij\rangle}\sum_{\lambda}\sum_{\sigma} K_{\lambda\sigma}\mathcal{P}_{\lambda}(S_i)\mathcal{P}_{\sigma}(S_j)-\sum_{i}\sum_{\lambda}\mu_{\lambda}\mathcal{P}_{\lambda}(S_i)
	\label{ternary H}
\end{equation}

where $\lambda,\sigma=1,2,3$ label the different particles, $\mu_{\lambda}$ are the chemical potentials and the $\mathcal{P}_{\lambda}$ are the projection functions which we explain now. The projection functions have the property that
\begin{equation}
	\sum_{\lambda=0}^{3}\mathcal{P}_{\lambda}(S)=1\hspace{1cm};\hspace{1cm}\mathcal{P}_{\lambda}(S)\mathcal{P}_{\nu}(S)=\delta_{\lambda\nu}\mathcal{P}_{\nu}(S)
	\label{projec1}
\end{equation}
and are easy to construct as polynomials of third order in the spin variable $f(S)= a_0+a_1\,S+a_2\,S^2+a_3\,S^3$ such that, for example, $\mathcal{P}_{1}(S)=1$ for $S=3/2$ and zero otherwise, etc. The projection functions turn out to be 
\begin{eqnarray}
	\mathcal{P}_{1}(S)&=& \frac{1}{48} \left(8 S^3+12 S^2-2 S-3\right)\nonumber\\
	\mathcal{P}_{2}(S)&=& \frac{1}{16} \left(-8 S^3-4 S^2+18 S+9\right)\nonumber\\
	\mathcal{P}_{3}(S)&=& \frac{1}{16} \left(8 S^3-4 S^2-18 S+9\right) \nonumber\\
	\mathcal{P}_{4}(S)&=& \frac{1}{48} \left(-8 S^3+12 S^2+2 S-3\right). 
	\label{projec2}
\end{eqnarray}

Plugging the projection operators back in the Hamiltonian, eq.(\ref{ternary H}), and rearranging the spin terms one obtains the most general nearest neighbour spin-3/2 Hamiltonian. We present its one-dimensional version,

\begin{eqnarray}
	\mathcal{H}&=&-J\sum_{i}S_i\,S_{i+1}-K\sum_{i}S_i^2\,S_{i+1}^2-L\sum_{i}S_i^3\,S_{i+1}^3\nonumber\\
	&-&  \frac{M_1}{2}\sum_{i}(S_i^2\,S_{i+1}+S_i\,S_{i+1}^2)-\frac{M_2}{2}\sum_{i}(S_i\,S_{i+1}^3+S_i^3\,S_{i+1})\nonumber\\
	&-& \frac{M_3}{2}\sum_{i}(S_i^2\,S_{i+1}^3+S_i^3\,S_{i+1}^2)-H\sum_{i}S_i+D\sum_{i} S_i^2\nonumber\\
	&-& W\sum_{i}S_i^3.
	\label{spin3by2}           
\end{eqnarray}

 The spin-3/2 case admits three order parameters, namely the mean magnetization per site $M$, the mean quadrupole moment per site $Q$ and the mean octopole moment per site $\Omega$.
 
 \begin{eqnarray}
 	M &=& \langle S_i\rangle\, , \nonumber\\
 	Q &=& \langle S_i^2\rangle\,\,\,\,\,\mbox{and}\nonumber\\
\Omega &=&\langle S_i^3\rangle\,.
\label{order parameters}
 \end{eqnarray}
We note that owing to translation invariance the  lattice index subscript $i$ on the spin variable $S$ is of no consequence.

Each of the nine spin coupling strengths $J,K,..,W$ in eq.(\ref{spin3by2}) above are given in terms of the nine lattice gas couplings, namely the six $K$'s and the three $\mu$'s. The spin coupling strengths $H,D\, \mbox{and}\, W$ depend on all the six $K$'s and the three $\mu$'s as can be checked. Considering the potentials $\mu$ as external, tunable parameters we can reinterpret the couplings $H$, $D$ and $W$ as external, tunable `fields'. On the other hand the remaining six spin couplings $J,K,L,M_1,M_2$ and $M_3$ depend only on the six $K$'s and can be thought of as `internal'.

To keep things simple and symmetric we switch off the `cross' interactions as well as the octopole-octopole coupling strength $L$ to obtain the working Hamiltonian for this investigation,

\begin{eqnarray}
	\mathcal{H}_{3/2}&=&-J\sum_{i}S_i\,S_{i+1}-K\sum_{i}S_i^2\,S_{i+1}^2\nonumber\\
	&-& H\sum_{i}S_i+D\sum_{i} S_i^2-W\sum_{i}S_i^3.
	\label{Ham spin3by2}           
\end{eqnarray}

While the above Hamiltonian is drastically simplified, it is complex enough to include non-trivial effects peculiar to spin-3/2 lattice\footnote{Since the spin-spin coupling $J$ is always positive in this work, in all the calculations we shall always scale it away, though it will occasionally appear in the formulae in the usual way.}. The transfer matrix for the above Hamiltonian can be solved numerically to obtain the largest eigenvalue and the correlation function (more about it later). Setting $H$ and $W$ to zero renders additional symmetry to the transfer matrix and it becomes amenable to closed form solutions. Of course, now the model becomes a spin-3/2 BEG model as treated in \cite{sivad} in the context of $\mbox{DyVO}_4$ phase structure. But, as we shall see in the sequel, its geometry and phase structure already reflect the complexity of spin-3/2. 

 The transfer matrix for the zero field spin-3/2 BEG model obtained by setting $H=W=0$ in the Hamiltonian in eq.(\ref{Ham spin3by2}) is 

\begin{widetext}
	\begin{equation}
		T=\left(
		\begin{array}{cccc}
			e^{\frac{81 K \beta }{16}-\frac{9 D \beta }{4}+\frac{9 \beta }{4}} & e^{\frac{9 K \beta }{16}-\frac{5 D \beta }{4}+\frac{3 \beta }{4}} & e^{\frac{9 K \beta }{16}-\frac{5 D \beta }{4}-\frac{3 \beta }{4}} & e^{\frac{81 K \beta }{16}-\frac{9 D \beta }{4}-\frac{9 \beta }{4}} \\
			e^{\frac{9 K \beta }{16}-\frac{5 D \beta }{4}+\frac{3 \beta }{4}} & e^{\frac{K \beta }{16}-\frac{D \beta }{4}+\frac{\beta }{4}} & e^{\frac{K \beta }{16}-\frac{D \beta }{4}-\frac{\beta }{4}} & e^{\frac{9 K \beta }{16}-\frac{5 D \beta }{4}-\frac{3 \beta }{4}} \\
			e^{\frac{9 K \beta }{16}-\frac{5 D \beta }{4}-\frac{3 \beta }{4}} & e^{\frac{K \beta }{16}-\frac{D \beta }{4}-\frac{\beta }{4}} & e^{\frac{K \beta }{16}-\frac{D \beta }{4}+\frac{\beta }{4}} & e^{\frac{9 K \beta }{16}-\frac{5 D \beta }{4}+\frac{3 \beta }{4}} \\
			e^{\frac{81 K \beta }{16}-\frac{9 D \beta }{4}-\frac{9 \beta }{4}} & e^{\frac{9 K \beta }{16}-\frac{5 D \beta }{4}-\frac{3 \beta }{4}} & e^{\frac{9 K \beta }{16}-\frac{5 D \beta }{4}+\frac{3 \beta }{4}} & e^{\frac{81 K \beta }{16}-\frac{9 D \beta }{4}+\frac{9 \beta }{4}} \\
		\end{array}
		\right)
		\label{transfer 3by2}
	\end{equation}
\end{widetext}
This allows for a closed form expression of the largest eigenvalue,
 \begin{eqnarray}
	\lambda_+ &=&\frac{1}{2} e^{-\frac{9 \beta }{4}-\frac{9 \beta  D}{4}+\frac{\beta  K}{16}}
	\Big(e^{\frac{5 \beta }{2}+2 \beta  D}\nonumber\\
	&+&e^{2 \beta +2 \beta  D}+e^{5 \beta  K}+e^{\frac{9 \beta }{2}+5 \beta  K}+\sqrt{X^2-4 Y}\Big),
	\label{largeeig}
\end{eqnarray}
where
\begin{equation}
	X=-e^{2 \beta +2 \beta  D}-e^{\frac{5 \beta }{2}+2 \beta  D}-e^{5 \beta  K}-e^{\frac{9 \beta }{2}+5 \beta  K}\nonumber
\end{equation}
and
\begin{eqnarray}
	Y&=&-e^{3 \beta +2 \beta  D+\beta  K}-2 e^{\frac{9 \beta }{2}+2 \beta  D+\beta  K}-e^{6 \beta +2 \beta  D+\beta  K}\nonumber\\&+&e^{2 \beta +2 \beta  D+5 \beta  K}	+e^{\frac{5 \beta }{2}+2 \beta  D+5 \beta  K}+e^{\frac{13 \beta }{2}+2 \beta  D+5 \beta  K}\nonumber\\&+&e^{7 \beta +2 \beta  D+5 \beta  K}.\nonumber
\end{eqnarray}
The Massieu function per spin (free energy, in short) can be obtained as the log of the largest eigenvalue $\lambda_+$,
\begin{equation}
	\psi = \log \lambda_+
\label{free energy log}	
\end{equation}	
More generally, we will retain a non-zero $H$ and $W$ so that the transfer matrix can now be solved only numerically implying that the free energy and subsequent operations on it are performed numerically.

The correlation function between spins $R$ lattice sites apart is

\begin{equation}
	\langle S^{\alpha}_{1}S^{\beta}_{1+R}\rangle -	\langle S^{\alpha}_{1}\rangle \langle S^{\beta}_{1+R}\rangle =\sum_{j \neq 1}\left(\frac{\lambda_j}{\lambda_1}\right)^R\,\langle t_j|{\bf{S}^\alpha}|t_1\rangle\,\langle t_1|{\bf {S}^\beta}|t_j\rangle,
	\label{correlation}
\end{equation}
where ${\bf{S}}$ is the spin matrix \footnote{Similarly, the quadrupole and octopole matrices are ${\bf{S^2}}$ and 	${\bf{S^3}}$.}
\begin{equation}
	{\bf{S}} = \left(\begin{array}{cccc}
		\frac{3}{2}&0&0&0\\0&\frac{1}{2}&0&0\\0&0&-\frac{1}{2}&0\\0&0&0&-\frac{3}{2}\\
	\end{array}\right)
	\label{S matrix}
\end{equation}
 Here $|t_j\rangle$ denote the eigenvectors of the transfer matrix so that we have $T=\sum_i |t_i\rangle \lambda_i|\langle t_i|$ with eigenvalues $\lambda_+=\lambda_1>\lambda_2>\lambda_3>\lambda_4$.
For non-zero $H$ or $W$, it can be checked that the matrix elements $\langle t_2|{\bf {S}^{\alpha}}|t_1\rangle$  are non-zero for $\alpha =1,2,3$ so that in this case there is only one correlation length 
\begin{equation}
	\xi_m^{-1}=-\log{\left|\frac{\lambda_2}{\lambda_1}\right|}.
	\label{xim}
\end{equation}

However, when both $H$ and $W$ are set to zero the matrix element  $\langle t_2|{\bf {S}^{2}}|t_1\rangle$ becomes zero with the leading non-zero term element being $\langle t_3|{\bf {S}^{2}}|t_1\rangle$. Thus, in this case there are two correlation lengths, the above one for the dipole moment and
\begin{equation}
	\xi_{q}^{-1}=-\log{\left|\frac{\lambda_3}{\lambda_1}\right|}.
	\label{xiq}
\end{equation} 

for the quadrupole moment. On the other hand, the correlation length for the octopolar fluctuations is always same as the dipolar correlation length, namely $\xi_m$ in eq.(\ref{xim}). This is because for the octopolar case the matrix element  $\langle t_2|{\bf {S}^{3}}|t_1\rangle$ is already non-zero everywhere. Satisfyingly, as we shall show in the sequel, geometry robustly encodes this feature.

We also point out an interesting feature of the quadrupole fluctuations. Using eq. (\ref{correlation}) it can be checked that for low temperatures the second moment of quadrupole fluctuations per lattice site $\sigma_q^2$ goes as
\begin{equation}
	\sigma_{q}^{2}=\frac{|\langle t_1|{\bf{S}^2}|t_3\rangle|^2}{1-\lambda_3/\lambda_1}\sim \frac{|{Q-Q_0}|}{1-\lambda_3/\lambda_1}	.
	\label{qsigma}
\end{equation} 
This expression is similar to the spin-one case in  \cite{krinsky}.

In the cases for which $\xi_q$ diverges, the denominator of eq.(\ref{qsigma}) goes to zero. However, depending on the {\textit{relative speed}} with which the numerator and denominator tend towards zero , $\sigma_q^2$ might or might not diverge. In particular, there will also occur counter-intuitive cases where $\sigma_q^2$ decays even as the correlation length $\xi_q$ diverges. In the special case where both the numerator and denominator decay equally fast,  $\sigma_q^2$ would approach a constant value at low temperature. Remarkably, as we shall demonstrate in the sequel, geometry will efficiently encode all these cases.

 
\section{Ground state phase structure of the spin-3/2 chain }
\label{phastru}


\begin{figure*}[t!]
	\begin{subfigure}[b]{0.3\textwidth}
		\centering
		\includegraphics[width=2.3in,height=1.8in]{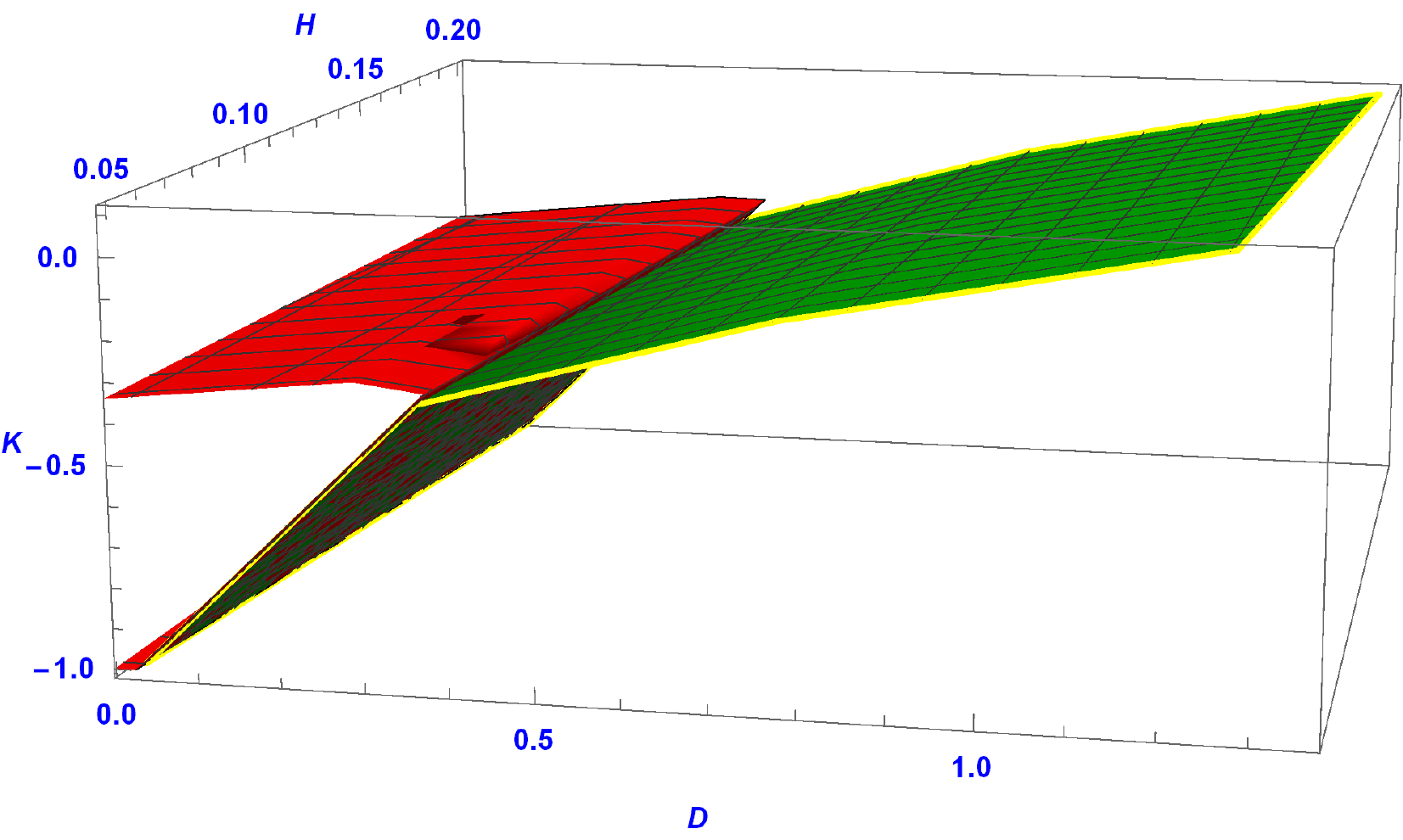}
		
		\caption{}
	\end{subfigure}
	\hspace{1.4in}
	\begin{subfigure}[b]{0.4\textwidth}
		\centering
		\includegraphics[width=2.3in,height=1.8in]{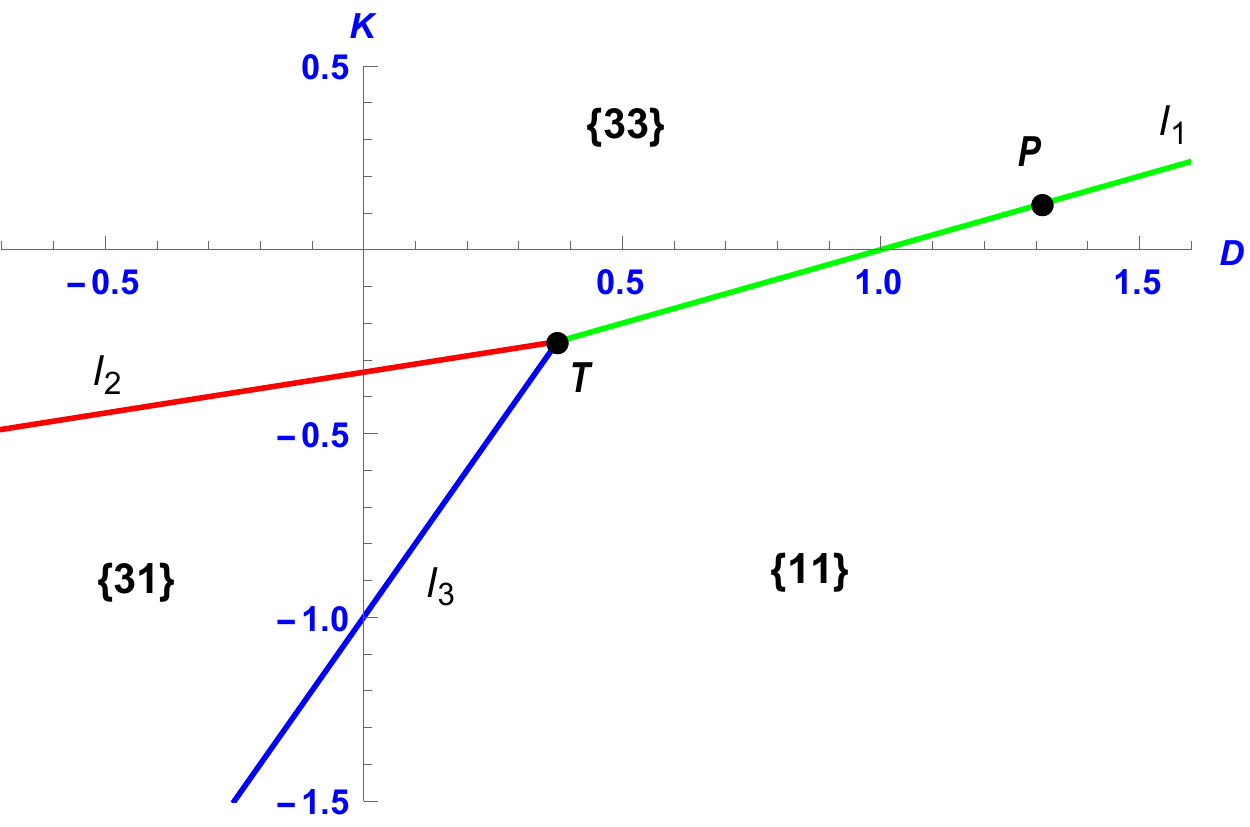}
		
		\caption{}
	\end{subfigure}
	\caption{\small{Ground state phase diagram for the spin- 3/2 model in ($a$) $D$-$K$-$H$ space with $W = 0$. The three coexistence planes partitions the space into three phases. The triple line is the intersection of these three planes. ($b$) The projection of ($a$) in $K$-$D$ plane with $H = 0$. Here the projection of triple line is point T. }}
	\label{phasekd1}
\end{figure*}

\begin{figure*}[t!]
	\begin{subfigure}[b]{0.3\textwidth}
		\centering
		\includegraphics[width=2.3in,height=1.8in]{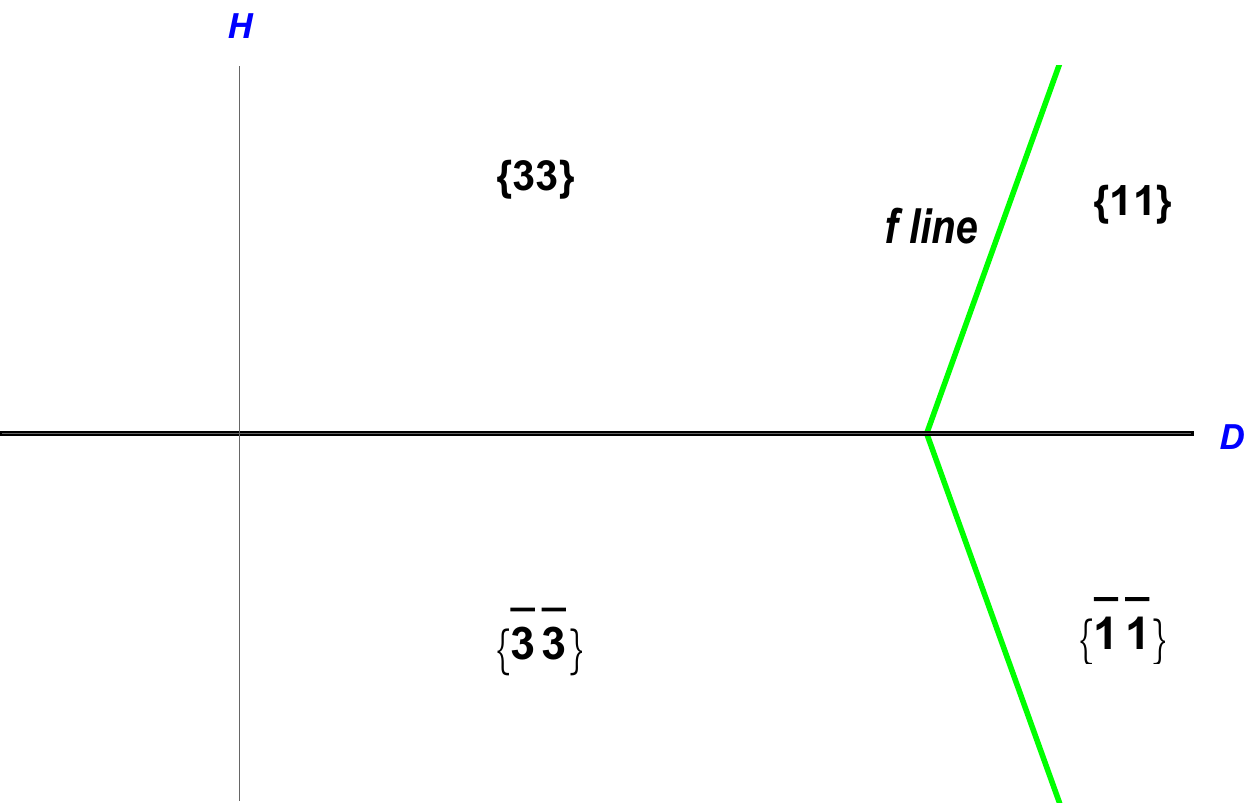}
		
		\caption{}
	\end{subfigure}
	\hspace{1.4in}
	\begin{subfigure}[b]{0.4\textwidth}
		\centering
		\includegraphics[width=2.3in,height=1.8in]{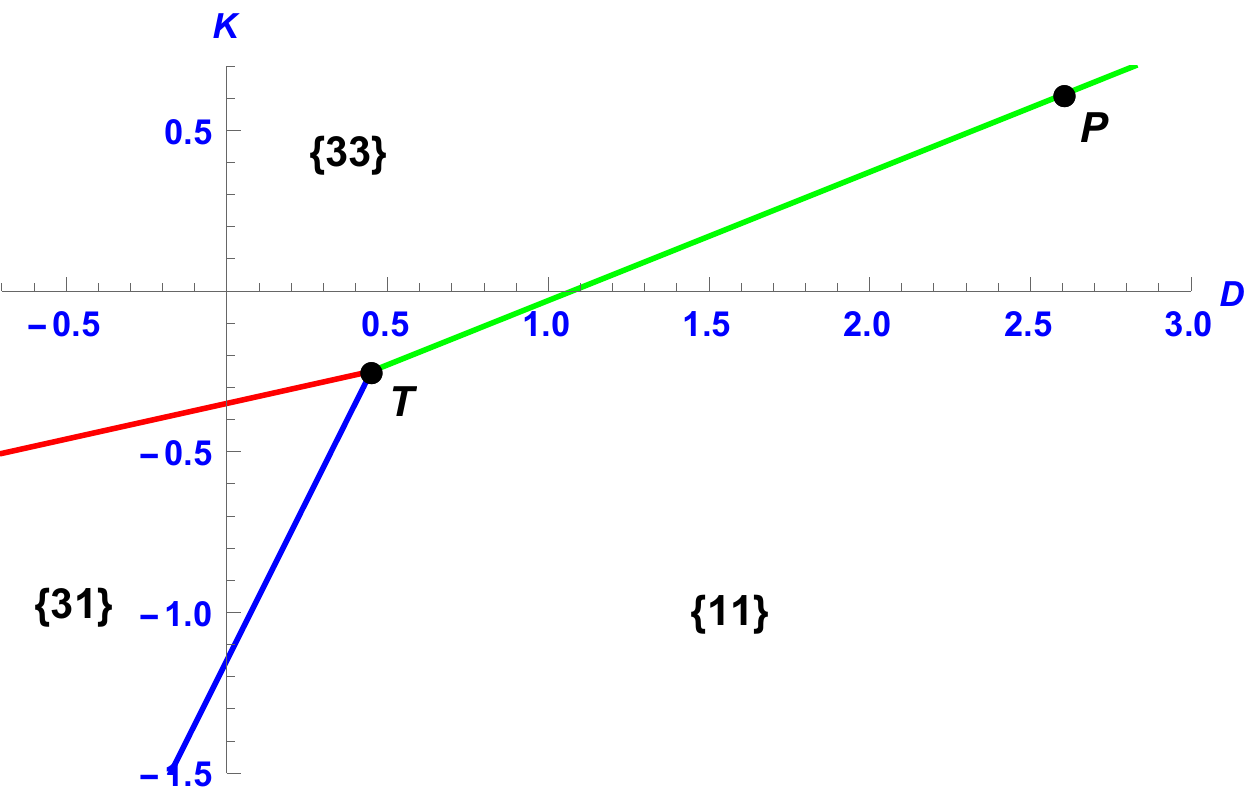}
		
		\caption{}
	\end{subfigure}
	\caption{\small{Ground state phase diagram for the spin-3/2 model ($a$) in the $H$-$D$ plane with  $W=0, K=0.1$ and ($b$) the non-zero field `$l_1$', `$l_2$' and `$l_3$' lines in the $D$-$K$ plane with $H=0.15$. The point ${\bf{P}}$ beyond which both dipole and quadrupole fluctuations decay is at $D=2.6$ and $K=0.6$. Compare with the zero-field $l_1$ line in fig.(\ref{phasekd1}b).  }}
	\label{gs}
\end{figure*}

The zero temperature phase structure is realized by comparing the  energies of different ground state configurations obtained from the spin-3/2 Hamiltonian in eq.(\ref{Ham spin3by2}). Thus one obtains the ground state energy of a nearest-neighbour pair with spins $S$ and $S'$, \cite{barreto}
\begin{eqnarray}
	\mathcal{E}_{SS'}&=& -J\,S\,S'-K\,S^2\,S'^2-\frac{H}{2}(S+S')+\nonumber\\
	&&\frac{D}{2}(S^2+S'^2)-\frac{W}{2}(S^3+S'^3)
	\label{energy eq}
\end{eqnarray}
which may be minimized over different nearest-neighbour configurations conveniently labeled as $\{33\},\{\bar{3}\bar{3}\},\{31\},\{\bar{3}\bar{1}\}, \{11\},\{\bar{1}\bar{1}\}$, etc. 
Here, for example $\{3\bar{3}\}$ would represent the nearest-neighbour configuration $\{\frac{3}{2},-\frac{3}{2}\}$, etc.
Since the Hamiltonian is reflection symmetric a change in sign of $H$ and $W$ together will
simply flip the sign of the stable spin configurations so that we will not lose any generality by considering only the cases of $H,W$ both positive and the case of $H$ positive and $W$ negative. 
For ferromagnetic $J$ (which is always the case in our investigation) the opposite sign pairs like 
$\{3\bar{1}\}$, etc are always of a higher energy so they will not figure anywhere in the phase diagrams to follow.

 For the case $H,W$ both non-negative we obtain from the pair-energy equation eq.(\ref{energy eq}) three planes of coexistence separating the $H-D-K$ space, parametrized by $W$,  into three ground-state configurations  ${\{33\}}$, ${\{11\}}$, and ${\{31\}}$. Thus, the planes $\mathcal{L}_1$, $\mathcal{L}_2$, and $\mathcal{L}_3$ given as
\begin{eqnarray}
	8D  - 20 K-4H &=&8 + 13 W \nonumber\\
	8 D-36 K-4 H&=&12 +13 W \,\,\mbox{and}\nonumber\\
	8 D  -4 K-4H&=&4 + 13 W
\label{coex planes}
\end{eqnarray}
separate, respectively, ${{\{33\}}}$ and ${{\{11\}}}, {{\{33\}}}$ and ${{\{31\}}}$, and finally, ${{\{31\}}}$ and ${{\{11\}}}$ ground state configurations. As we shall see subsequently, while the coexistence plane $\mathcal{L}_1$ shows criticality even for non-zero $H$ and $W$ the $\mathcal{L}_2, \mathcal{L}_3$ planes become critical only when they intersect with the plane $H=W=0$ in what we shall term as the $l_2$ and $l_3$ lines in the following.

The coexistence planes intersect in the triple line in the $H-D-K$ space, given by the equations
\begin{equation}
   2D-H=\frac{3}{4} +{\frac{13W}{4}}\,\,\,\,\mbox{and}\,\,\,\,K=-\frac{1}{4}.
\label{triple line}
\end{equation}

We shall discuss separately the phase structures of the case $W=0$, namely the BEG spipn-3/2 case, and the more general case of non-zero $W$. The former, restricted case will be investigated more comprehensively while in the latter, more general case we shall not claim any completeness. 

 \subsubsection{\bf{Phase behaviour in the BEG case, $W=0$}}
 \label{gs1}

 Fig.(\ref{phasekd1}a) shows the phase diagram in the $H-D-K$ parameter space of the spin-3/2 BEG chain, namely the Hamiltonian in eq.(\ref{Ham spin3by2}) with $W$ set to zero. Above the green coloured plane $\mathcal{L}_1$  extending to the top right and the red colored plane $\mathcal{L}_2$ extending to the left, ${\{33\}}$ remains the ground state of the chain. Below the plane $\mathcal{L}_1$ the configuration ${\{11\}}$ becomes the ground state. Finally, sandwiched between $\mathcal{L}_2$ and the black colored plane $\mathcal{L}_3$ extending to the bottom left, the  ${\{31\}}$ configuration becomes the most stable. Henceforth, we shall label the region with globally stable ${\{33\}}$ ground state configuration as the ${\bf{{\{33\}}}}$ region, etc.
 
 Fig.(\ref{phasekd1}b) is a zero-field projection of the phase diagram of fig.(\ref{phasekd1}a) on the $D-K$ plane with $H=W=0$. The lines labelled $l_1$, $l_2$ and $l_3$ represent the intersections of the corresponding coexistence planes with the plane $H=0$ in the $H-D-K$ space. The three coexistence lines intersect at the triple point ${\bf{T}}$ with $\{D=3/8,K=-1/4\}$. This phase diagram was first reported in \cite{barreto}.


\begin{figure*}[t!]
	\begin{subfigure}[b]{0.3\textwidth}
		\centering
		\includegraphics[width=2.8in,height=2.1in]{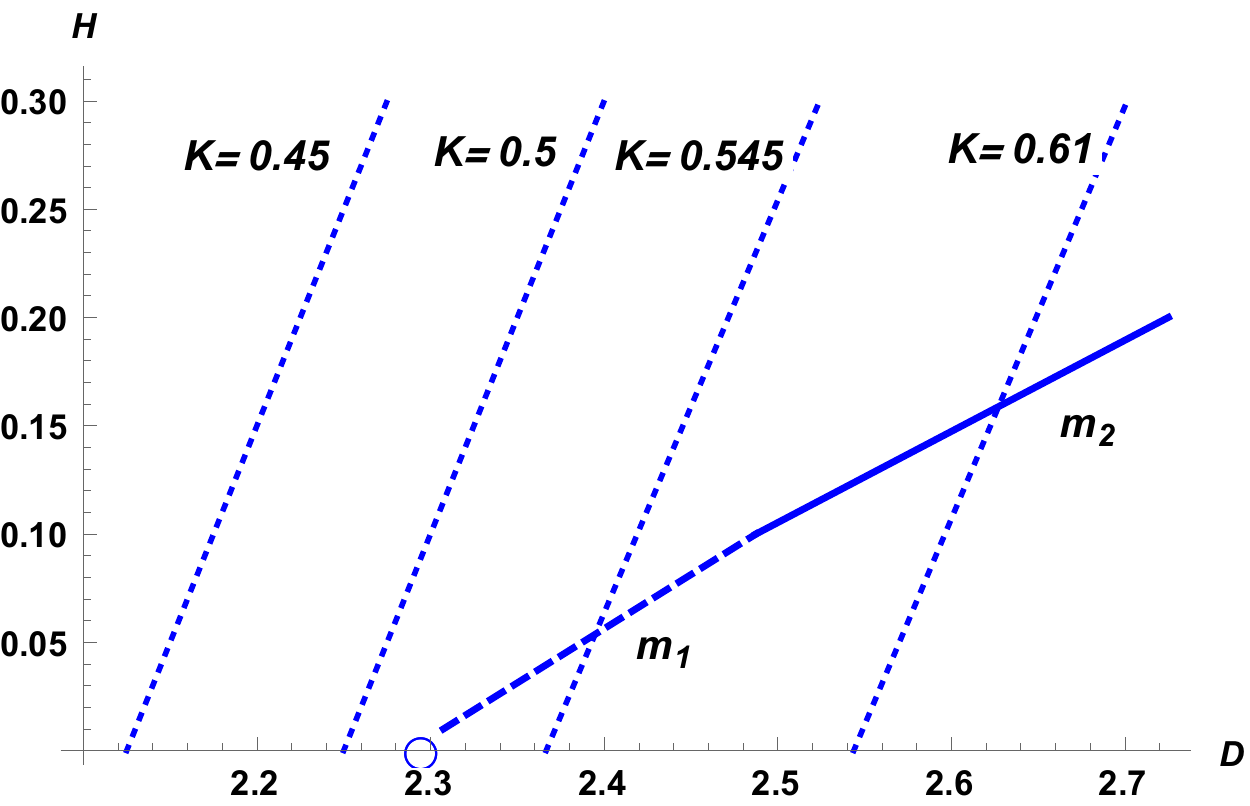}
		
		\caption{}
	\end{subfigure}
	\hspace{1.4in}
	\begin{subfigure}[b]{0.4\textwidth}
		\centering
		\includegraphics[width=2.8in,height=2.1in]{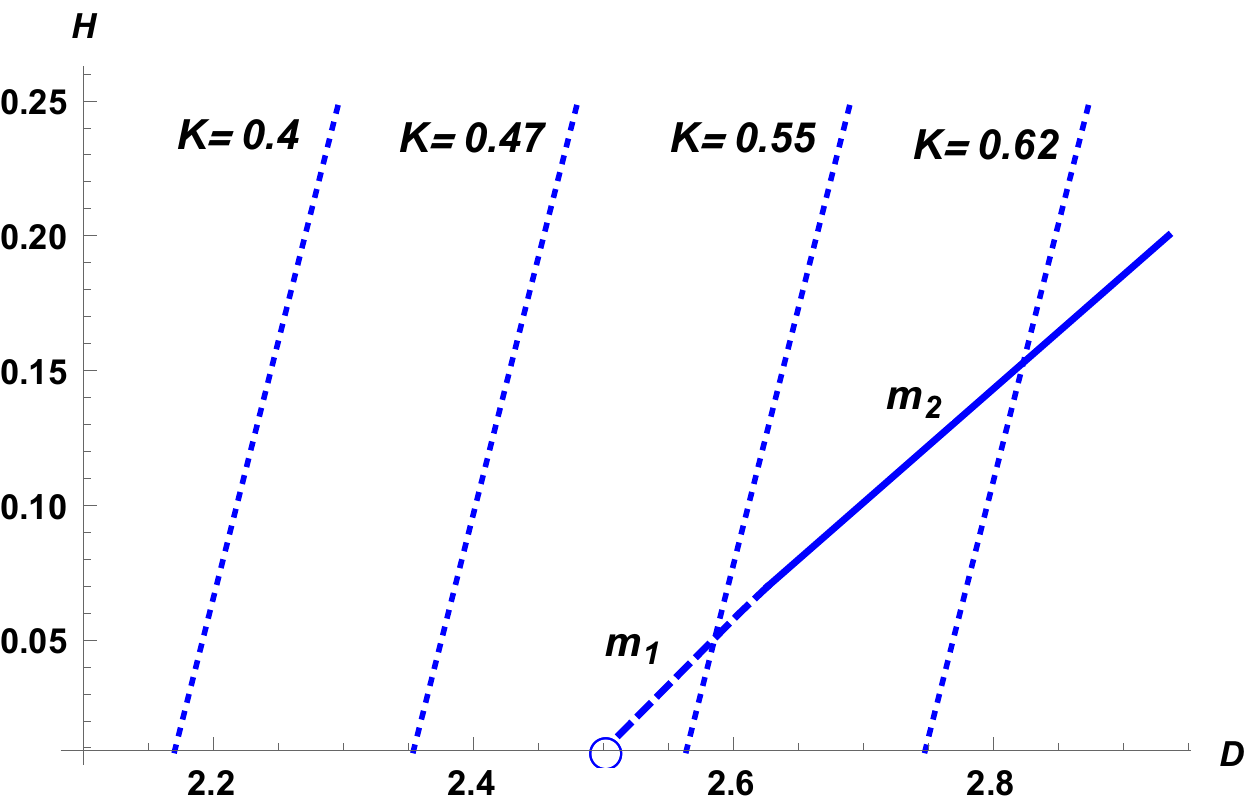}
		
		\caption{}
	\end{subfigure}
	\caption{\small{($a$) Projection of the $\mathcal{L}_1$ coexistence plane onto the $D-H$ plane for the spin-3/2 BEG chain, $W=0$. The $m_1$ and $m_2$ lines separate regions of diverging fluctuations to their left and decaying fluctuation moments to their right. The dotted lines are the $f$ lines like the one in fig.(\ref{gs}b) with different $K$ values. ($b$) Similar to ($a$) but for the general spin-3/2 chain with $W=1/10$. Open circles on the $H$ axis in both sub-figures indicate that the $m$ lines are defined only for non-zero $H$. The ${\bf{P}}$ point at $H=0$ is discontinuous with other ${\bf{P}}$ points.}}
	\label{khatualines}
\end{figure*}

 The whole of the zero field  phase diagram ($H=W=0$) of the BEG case is critical for the spin fluctuations and octopole fluctuations. However, it can be checked, except for a segment of the coexistence line $l_1$, the quadrupole fluctations are finite everywhere and mostly decay to zero for low temperatures. Starting with the triple point ${\bf{T}}$ at $D=3/8$, where the variance $\sigma_q^2$ approaches unity, the quadrupole fluctuations show a slow divergence to infinity which becomes steeper along $l_1$ until the point $D=1,K=0$ where the divergence is the sharpest. Moving further to the right where $K$ is positive, the divergence slows down until $\sigma_q^2$ flattens to $8$ at the  point $\{D=21/16, K=1/8\}$ labeled as $\bf{P}$ in fig.(\ref{phasekd1}b). Beyond $\bf{P}$ the quadrupole fluctuations decay to zero on the line $l_1$. Finally, the moment $\sigma_q^2$ always approaches small positive values on both lines $l_2$ and $l_3$ while everywhere else in the zero field plane it decays to zero as mentioned earlier.
 
Moving on to non-zero values of $H$ we note that the coexistence planes $\mathcal{L}_2$ and $\mathcal{L}_3$ of fig.(\ref{phasekd1}a)  remain non-critical with a decaying correlation length\footnote{We recall that for non-zero $H$ or $W$ there is only one correlation length ($\xi_m$ of eq.(\ref{xim})) for all the order parameters.} while the coexistence plane $\mathcal{L}_3$ shows criticality, with the correlation length diverging everywhere on it\footnote{ Indeed, the order parameter fluctuation do not diverge everywhere on $\mathcal{L}_1$ as we shall soon see.}.

We study the non-zero field phase behaviour of the spin-3/2 BEG case in fig.(\ref{gs}a) which shows the intersection of the coexistence planes of fig. (\ref{phasekd1}a) with a fixed $K$ plane for a representative value of $K$ above the triple line at $K=-1/4$. For $K=0.1$ there is only one phase boundary, the $f$ line corresponding to the $\mathcal{L}_1$ plane as shown in fig.(\ref{gs}a) where the correlation length diverges everywhere. In fig.(\ref{gs}b) we depict the intersection of the phase coexistence surfaces of fig.(\ref{phasekd1}a) withe the surface $H=0.15$. The coexistence lines are similar to the lines $l_1$, $l_2$ and $l_3$ of fig.(\ref{phasekd1}b) however now the non-zero field lines $l_2$ and $l_3$ are completely non-critical, while the line $l_1$ shows critical fluctuations in order parameters upto the point ${\bf{P}}$ which is $H$ dependent (and whose equation we shall soon  discuss). Beyond the point ${\bf{P}}$ the fluctuations decay, though the correlation length still diverges. 

\begin{figure*}[t!]
	\begin{subfigure}[b]{0.3\textwidth}
		\centering
		\includegraphics[width=2.3in,height=1.8in]{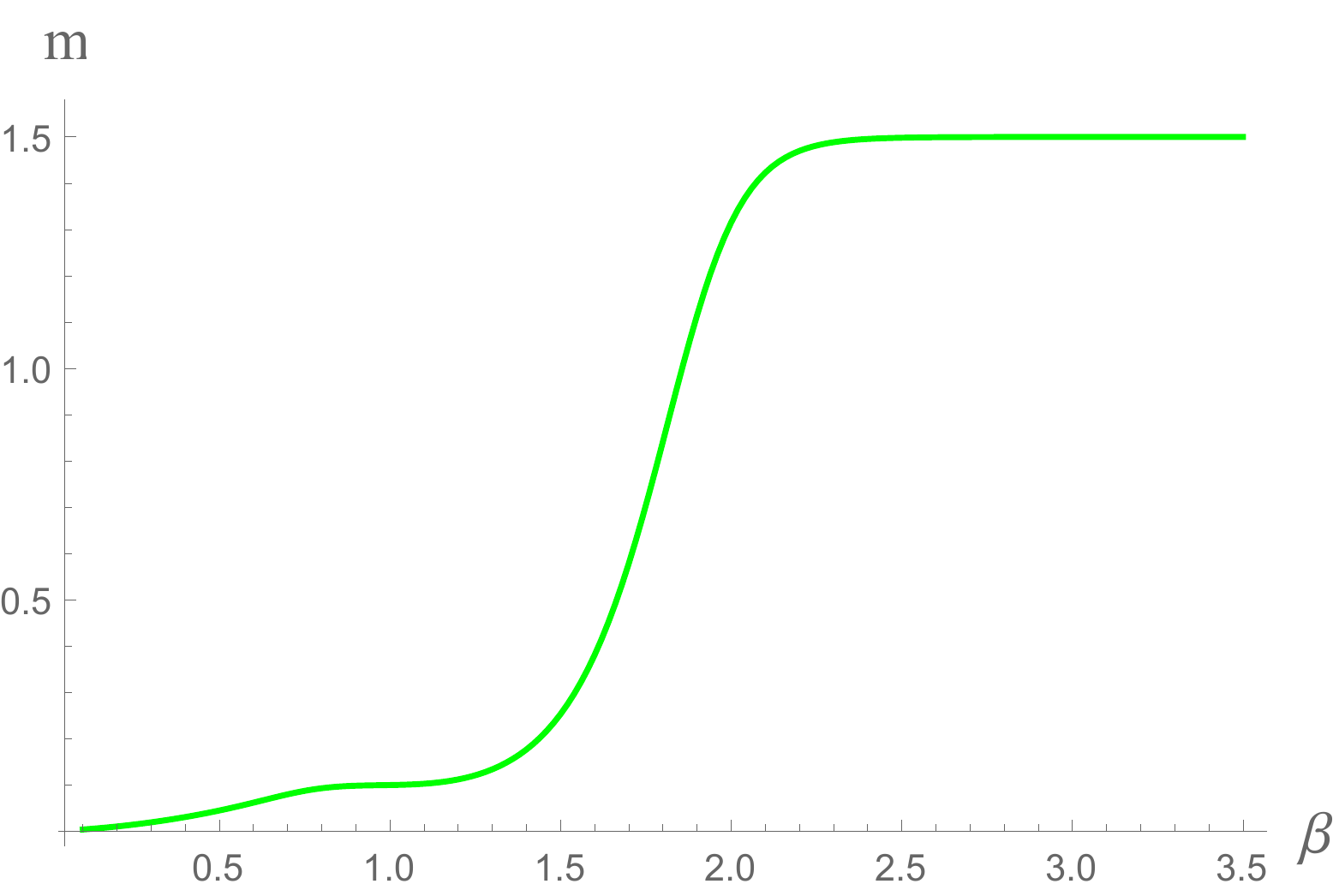}
		
		\caption{}
	\end{subfigure}
	\hspace{1.4in}
	\begin{subfigure}[b]{0.4\textwidth}
		\centering
		\includegraphics[width=2.3in,height=1.8in]{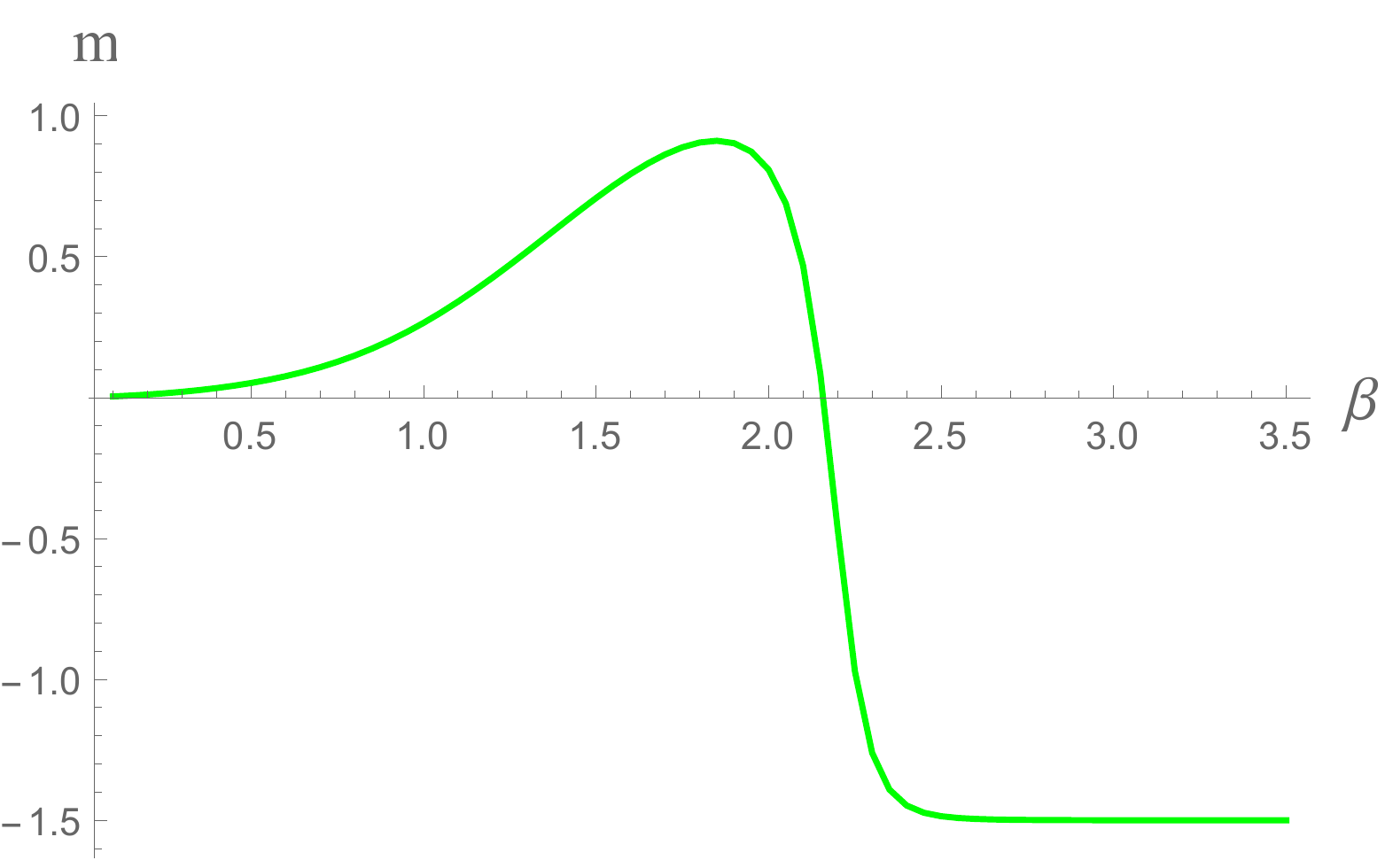}
		
		\caption{}
	\end{subfigure}
	\caption{\small{Magnetization $m$ vs $\beta$ plots for positive $H$ and negative $W$. In (a) where $H$ dominates $W$ the spin sharply increases to $3/2$ from small values. The parameters are $H=0.3$, $K=1$, $D=3$, $W=-0.1333$. In (b) where $W$ just about dominates $H$ spin flips sharply to $-3/2$ starting from positive values at high temperature. The parameters are $H=0.3$, $K=1/8$, $D=6/8$, $W=-0.1335$. }}
	\label{negative W}
\end{figure*}

We therefore see that the plane $\mathcal{L}_1$ is not critical everywhere. It shows further interesting structure which, we believe, is presented for the first time here. Thus, in fig.(\ref{khatualines}a), where the coexistence plane $\mathcal{L}_1$ has been projected onto the $H-D$ plane, the numerically obtained straight lines $m_1$ (blue, dashed) and $m_2$ (blue, smooth) divide the plane into a critical left part where the quadrupole, dipole and octopole fluctuations all diverge and a non-critical right part where they decay towards zero. These $m$ lines are nothing but a locus of the points ${\bf{P}}$ for different values of $H$. On the lines themselves all the fluctuations flatten to small finite values. For example, $\sigma_m^2\to 2$ everywhere on the line $m_2$ while on the line $m_1$ it flattens to decreasing values starting from $2$ as $H$ approaches zero. The equation for the $m_1$ and $m_2$ lines for $W=0$ is obtained by a numerical investigation as
\begin{eqnarray}
80D &=& 160 H+183\hspace{0.36in}m_1\text{ line, }W=0\nonumber\\
8D &=& 19 H+18\hspace{0.5in}m_2\text{ line, }W=0
\label{m1m2line}
\end{eqnarray}

  The dotted lines in fig.(\ref{khatualines}a) are the projections of fixed $K$ lines in the $\mathcal{L}_1$ plane onto the $H-D$ plane, thus being the $f$ lines of fig.(\ref{gs}a). The $f$ line labeled by $K= 0.5$ does not intersect the $m$ lines anywhere so that it is critical everywhere. On the other hand the $f$ lines marked $K=0.545$ and $K=1.23$ remain critical only to the left of the $m$ lines. From the equation for the $m_1$ line we can verify that for
\begin{equation}  
   -0.25<K<0.515
\label{kcritical} 
\end{equation}  
the $f$ line remains critical everywhere. 
Note that the line $m_1$ ends on the $H=0$ axis in an open circle indicating that it is defined only for non-zero $H$. Namely, the ${\bf{P}}$ point for zero magnetic field is discontinuous with such points for non-zero magnetic field.

 It is notable that irrespective of the $m$ lines, everywhere on the $\mathcal{L}_1$ plane the correlation length  $\xi$ diverges as mentioned earlier. A mathematical explanation of the curious fact that fluctuations decay to the left of the $m$ lines even as the correlation length diverges there is along similar lines as the discussion around eq.(\ref{qsigma}) for $q$-fluctuations. Thus, at low temperatures the spin and quadrupole fluctuation moments go respectively as,
\begin{eqnarray}
\sigma_{m}^{2}&\sim&\frac{|\langle t_1|{\bf{S}}|t_2\rangle|^2}{1-\lambda_2/\lambda_1},\nonumber\\
\text{and}\,\,\,\,\sigma_{q}^{2}&\sim&\frac{|\langle t_1|{\bf{S}^2}|t_2\rangle|^2}{1-\lambda_2/\lambda_1}.
\label{mlinemoments}
\end{eqnarray}
Once again, as with the $q$-fluctuations on the $l_1$ line, it can be checked that  to the left of the $m$ lines in fig.(\ref{khatualines}a)
the numerator approaches zero faster than the denominator.
 It will be demonstrated in the sequel that  the fluctuations and correlations in the dipolar and quadrupolar order parameters are efficiently encoded by geometry.

\subsubsection{\bf{Phase behaviour in the general case, $W\neq 0$}}
\label{gs2}

We now briefly report few instances of phase behaviour in the general spin-3/2 chain. As mentioned earlier, in the presence of a finite coupling $W$ the zero $H$ field plane in the $H-D-K$ space is no more a plane of symmetry and so we do not expect it to be critical everywhere, unlike the BEG case. Indeed, this holds true everywhere on the plane $H=0$  where the $m$ and $q$ fluctuations do not diverge, except, however, for a segment of the $l_1$ line from (but excluding) the triple point $\bf{T}$ to a $W$ dependent point $\bf{P}$ \footnote{Note that we are using the same labels for the $W\neq 0$ case coexistence lines and planes, triple point, etc as for the BEG case }. For example for $W=0.1$ and $W=0.5$ respectively, the $\bf{P}$ points are at $\{D,K\}$ equals $\{0.51,2.44\}$ and $\{0.65,3.44\}$. We also note that to the left of the $\bf{P}$ point the magnetization saturates to $1$ on the $l_1$ line and to its right it saturates to $0.5$. As in the previous cases, the correlation length diverges along the whole of the $l_1$ line even as the fluctuations decay beyond the $\bf{P}$ point.

For non-zero $H$ (same sign as $W$) the phase structure of the spin-3/2 chain remains qualitatively the same as the BEG case. Thus, the planes $\mathcal{L}_2$ and $\mathcal{L}_3$ are non-critical while $\mathcal{L}_1$ shows criticality, with the correlation length diverging everywhere on it. Once again, the $\mathcal{L}_1$ plane can be separated into a part where the fluctuations diverge and a part where the fluctuations decay.  From fig.(\ref{khatualines}b) for a general spin-3/2 chain with $W=1/10$ we find that the features on the critical surface $\mathcal{L}_1$ are qualitatively similar.

 A competition for the ground state arises when $H$ and $W$ are of opposite signs. This is because both the fields are coupled to odd powers of the lattice spins so that they have the same symmetry. Thus, for instance, when $H$ is positive and $W$ negative the former will compete for a positive spin oriented ground state as it will lower energy while the latter will prefer a negative spin ground state.  In one such case of competing configurations we just record our observation that for 
 \begin{equation}
 W<-\frac{4}{9}\,H
  \label{W negative}
 \end{equation}
 the $\{\bar 3\bar 3\}$ ground state becomes more stable than the $\{33\}$ state, irrespective of the values of $D$ and $K$. The latter do determine the sharpness with which the magnetization switches sign at a finite temperature, though we have not attempted any detailed investigation here. In fig.(\ref{negative W}a) and fig.(\ref{negative W}b) we show, respectively, magnetization vs $\beta$ plots for respectively positive $H$ dominant and negative $W$ dominant parameter values.  In subsequent sections we shall see that geometry will provide important clues to the underlying statistics.

\section{Method of constrained fluctuations: curvature invariants on hypersurfaces}
\label{threehgeo}

Thermodynamic geometry envisages a Riemannian manifold structure for the thermodynamic state space, which it achieves by introducing a non-negative distance measure between nearby equilibrium states, \cite{rupp1,rupporiginal,ruppfluc,weinhold} . In the entropy representation introduced and developed by Ruppeiner, \cite{rupporiginal} the thermodynamic metric can be conveniently represented as the second derivative of the Massieu function,
\begin{equation}
	g_{\mu\nu}=\frac{\partial^2\psi}{\partial x^{\mu}\partial x^{\nu}},
\end{equation}

where the co-ordinates $x^{\mu}$ are the entropic intensive variables. In a two dimensional state space, which is the case for simple fluids, Ising model, etc, the only independent curvature and hence the only source of microscopic information is the Riemann curvature scalar $R$. In three dimensions or higher the full Riemann curvature tensor $R_{\mu\nu\rho\sigma}$  comes into effect and it is worth exploring if different components of the Riemann tensor provide complementary information about the underlying microscopics. Alternatively, we could slice the higher dimensional state space via well-defined lower dimensional hypersurfaces and investigate the resulting  curvatures. The latter method, termed the method of constrained fluctuations, was adopted earlier in the context of the three-dimensional state space of the extended Kerr-Ads black holes in \cite{asknbads} and of the spin-one chain in \cite{riekan1}. 

The advantage of using carefully chosen sections is that a physically transparent meaning in terms of constrained fluctuations can be ascribed to the induced thermal metrics and thus also to the resulting lower dimensional  curvatures. Thus, in the spin-one case in \cite{riekan1} for example, we chose hypersurfaces of constant $H$ and of constant $D$ and were able to demonstrate that on the constant $H$ surface the fluctuations in magnetization $M$ were suppressed and those in the quadrupole $Q$ were unrestricted, while on the constant $D$ surface reverse was the case. In the former case of free $Q$ fluctuations the resulting sectional curvature was labelled $R_q$ and the latter sectional curvature was similarly labelled $R_m$. 

 In the present case, with a four-dimensional state space for the spin-3/2 chain, we can now construct three two-dimensional hypersurfaces by fixing any two of the three parameters $H, D\, \mbox{and}\, W$ \footnote{We have not pursued the full four-dimensional scalar curvature in this work. Apart from the high computational cost involved in its evaluation we feel it will contain mixed information about separate order parameter correlations and thus might not be very useful.}. Thus, the sectional curvatures on the hypersurface of fixed $D$ and $W$, called the $DW$-surface will be labelled $R_m$ as it might better encode the correlations in fluctuations of the dipole moment $M$ which are free on that surface, as against the quadrupole and the octopole fluctuations which remain somewhat suppressed \footnote{See, however, later discussion in this subsection on a three dimensional sclar curvature.}. Similarly, the sectional curvature on the $HW$-surface is labelled $R_q$ for encoding correlations in $Q$ fluctuations and the one on the $HD$-surface is labelled $R_{\omega}$ for encoding $\Omega$ correlations. It is important to note here that limiting thermodynamic fluctuations to the specified hypersurfaces is ultimately a mathematical device that helps filter out information about competing correlations. The only guiding principle is that the freezing out of fluctuations in certain directions should not be unphysical, as was clarified in \cite{riekan1}. On the other hand, it is also true that any physical process leading to a suppression or slowing down of certain fluctuations will automatically suggest a relevant hypersurface in the state space on which all thermodynamic motion (including fluctuations) remains restricted, as was the case in  \cite{asknbads}. 
 
 For mathematical details we refer the reader to \cite{asknbads} and \cite{riekan1} wherein a general formalism for choosing relevant hypersurfaces in a higher dimensional thermodynamic manifold was outlined. Also, the pullback of the ambient thermal metric on the chosen hypersurfaces was interpreted in terms of constrained fluctuations in the thermodynamic quantities. Here we shall simply outline the procedure for obtaining the three sectional curvatures starting with the free energy (Massieu function) per spin which is obtained as the logarithm of the partition function ,
 \begin{eqnarray}
  \psi=\frac{1}{N}\log \sum_{\{S_i\}}e^{ -\beta\, \mathcal{H}_{3/2}} . 
  \label{log partition}
 \end{eqnarray}
where $\mathcal{H}_{3/2}$ is the Hamiltonian in eq.(\ref{Ham spin3by2}). Considered as a function of $\beta$ with $H,D,W$ held constant, the beta derivative of the free energy will give the ``total'' energy $E$, which includes the ``internal'' energy $U$ of the lattice spins parameterized by the fixed couplings $J$ and $K$, and the ``external'' energy of coupling between the spins, quadrupoles and octopoles with the correponding tunable couplings $H$, $D$ and $W$,

\begin{equation}
	-\left.\frac{\partial{\psi}}{\partial\beta}\right|_{H,D,W}=E=\frac{1}{N}\langle{\mathcal{H}}_{3/2}\rangle
\end{equation}

On the other hand the internal energy $U$ can be obtained by taking the $\beta$ derivative of the free energy at a fixed value of the remaining entropic intensive variables $(\,\nu,\mu,\gamma\,)=(\,\beta\,H,\beta\,D,\,\beta\,W)$,

\begin{equation}
	-\left.\frac{\partial{\psi}}{\partial\beta}\right|_{\nu,\mu,\gamma}=U=E+HM-DQ+W\Omega
\end{equation}

 In terms of the entropic intensive variables the differential of free energy becomes,
 \begin{equation}
	d\psi=-U\,d\beta+ M\,d\nu-Q\,d\mu+\Omega\,d\gamma ,
	\label{dpsi1}
\end{equation}
By setting to constant two of the three parameters $H,D,W$  in the above equation we can obtain the governing equations for the three hypersurfaces. Thus, for the $DW$-surface eq.(\ref{dpsi1}) becomes

\begin{equation}
	d\psi_{DW}(\beta, \nu)=-(E+HM)\,d\beta+M\,d\nu ,
	\label{dpsiDW},
\end{equation}
and on the $HW$-surface the free energy differential becomes 
\begin{equation}
	d\psi_{HW}(\beta, \mu)=-(E-DQ)\,d\beta-Q\,d\mu ,
	\label{dpsiHW}
\end{equation}
while the $HD$ surface gives

\begin{equation}
	d\psi_{HD}(\beta, \gamma)=-(E+W\Omega)\,d\beta+\Omega\,d\gamma ,
	\label{dpsiHD}
\end{equation}

The calculations for the induced metrics and the sectional curvatures for the three hypersurfaces follow directly from the above three equations.

In addition to the sectional curvatures mentioned above, the $4$-dimensional Riemannian manifold also allows for the possibility of meaningful $3$-dimensional scalar curvatures. One such scalar curvature that we will explore in some detail is the one living on constant $D$ surfaces in the  $4$-dimensional  parameter space of the spin-3/2 chain. Restricting thermal fluctuations to within the constant $D$-hypersurface  will partially suppress the quadrupole fluctuations while allowing unrestricted variations in the spin and octopole moments. The latter two order parameters have a similar odd symmetry under sign change and hence their fluctuation statistics are also expected to be similar, given that they have the same correlation length as mentioned earlier. Therefore it might be useful to compare the $3d$ scalar curvature $R_D$ with the $2d$ scalar curvature $R_m$ discussed earlier.  Similar to the $R_D$ we could in principle also investigate other $3d$ scalar curvatures like $R_H$ and $R_W$ which live on state space sub-manifolds where, respectively, fluctuations in $H$ and $W$ are held frozen. However, since the $H$ and $W$ fields couple to the dipole and octopole moments which have a similar symmetry the respective curvatures $R_H$ and $R_W$ might contain mixed information about spin and octopole correlations on the one hand and the quadrupole correlation on the other. We shall therefore restrict our investigations to the $3d$ scalar curvature $R_D$.

\section{ scaling near pseudo-criticality}
\label{free}
We recall that the Ruppeiner equation relates the state space scalar curvature to the singular part of free energy (or the Massieu function $\psi_s$), \cite{rupporiginal},
\begin{equation}
R=\frac{\kappa_1}{\psi_s}
\label{rupp eq}
\end{equation}
where $\kappa_1$ is an order unity negative dimensionless constant whose value depends on the universal scaling exponents and the dimension of the state space manifold, \cite{rupp8}. Using hyperscaling, which relates the singular free energy to the correlation volume, in the asymptotic critical region we have
\begin{equation}
R=\kappa_2 \xi^d
\label{Rcorrel}
\end{equation}
where $d$ is the physical dimension of the system and $\kappa_2$ is similarly an order unity constant. However, as we shall verify for several cases here, the correspondence of $R$ with the correlation length $\xi$ continues in non-critical regimes as well even if it is not as mathematically precise as in eq.(\ref{Rcorrel}) near criticality. Thus, more generally,
\begin{equation}
R\sim\xi^d
\label{Rgeneral}
\end{equation}

This feature of $R$ has been successfully exploited in inferring the coexistence curves for fluid as well as magnetic spin systems, (cite widom line, mausbach, tapo, spin-one mean field).

  Before delving into the behaviour of the scalar curvatures in (pseudo)critical and non-critical regions we first summarize our findings for the scaling behaviour near (pseudo)criticality of the singular part of free energy followed by that of the correlation lengths.

\subsection{Scaling of the free energy}
\label{freeone}
 The scaling form of the free energy for the spin-3/2 chain near its pseudocritical points may be written as
\begin{align}
	\psi_{s}=n|\tau|^{p}Y\left(n_{1}\frac{h_1}{|\tau|^{q}} ,n_{2}\frac{h_2}{|\tau|^{r}},n_{3}\frac{h_3}{|\tau|^{s}}\right)
	\label{sin1}
\end{align}
where $p,q,r,s$ are the universal critical exponents and the $n$'s are non-universal constants. The scaling fields $h_1,h_2,\,\text{and}\,h_3$ could in general be obtained as linear combinations of the displacement of the fields $H,D\,\text{and}\,W$ from their critical values and $ \tau $ is the reduced temperature. $Y$ is the spin scaling function which becomes a constant when all the scaling fields are set to zero. In other words, at the (pseudo)critical values of $H,D\,\text{and}\,W$  the singular part of free energy depends only on the `reduced temperature' scaling field $\tau$. Note that, as in the spin-one case, \cite{riekan1}, the scaling field $\tau=e^{-\beta X}$ where $X$ can be a linear combination of the coupling constants $J,K$ and the fields.

In the case of zero field pseudocriticality of the BEG chain the free energy can be expressed in a closed form as already mentioned in the preceding. This allows us to analytically obtain the leading singular term in the free energy by filtering out the regular terms and the fast decaying terms, as in \cite{riekan1}.  
While for convenience we restrict ourselves in the following to the BEG case with $W=0$  we have checked that the free energy scaling  does not change for lines and regions which continue to remain critical in the general case $W\neq 0$. 

\begin{figure}[!h]
	\centering
	\includegraphics[width=3in,height=2.2in]{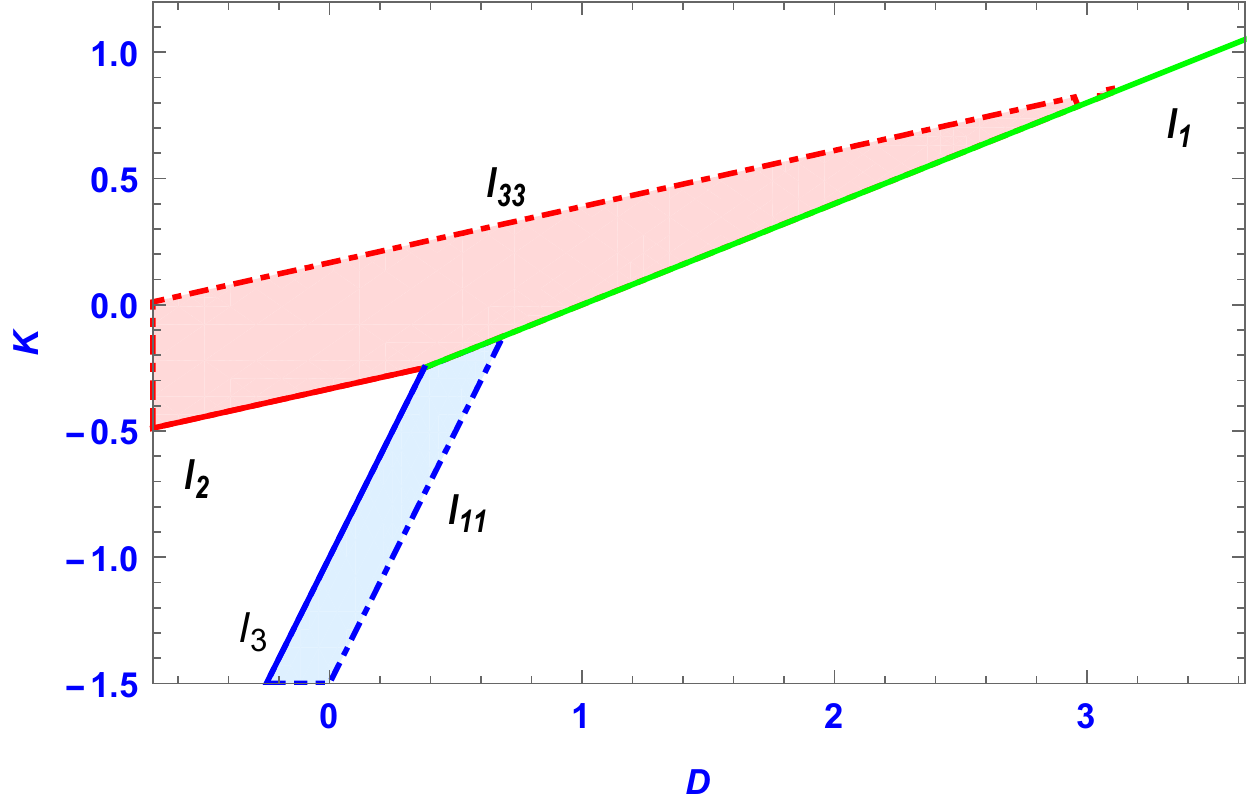}
	\caption{\small{ Ground state phase diagram for $W=0$ as in fig.(\ref{phasekd1}b) with additional lines $l_{33}$ and $l_{11}$ dividing respectively the regions {\{\bf{33}\}} and  {\{\bf{11}\}} into subregions with different scaling behaviours of the free energy. See text for details. }}
	\label{psiphase}
\end{figure}

We now present our results for the zero field scaling of the free energy for different regions and lines.
In the region $\bf{\{33\}}$ the singular free energy presents two types of scaling in sub-regions separated by the line $l_{33}$ given by $K=2 D/9+1/6$, see fig.(\ref{psiphase}). Note that $l_{33}$ is parallel to $l_2$ and shifted above $l_2$ by $J/2$. On and below the line $l_{33}$ and above it  the singular free energy has the following limiting expression in the region ${\bf{\{33}\}}$,
\begin{eqnarray}
	\psi_{s} &=& {e^{-(9K+3J-2D)\beta}}\,\,\,\,\,\text{below and on } l_{33},\nonumber\\
	\psi_{s} &=& {e^{-9\,J\beta/2}}\hspace{1in}\text{above } l_{33}.
	\label{psi4}
\end{eqnarray}
We shall label the light red shaded sub-region of ${\bf{\{33}\}}$ between $l_2$ and $l_3$ as ${\bf{\{33}\}'}$.

Similarly, in the region $\bf{\{11\}}$ we find a line $l_{11}$ (see fig.(\ref{psiphase}) parallel to $l_3$ defined by the equation $K = 2D-3J/2$ and shifted below $l_3$ by $J/2$ such that the limiting expression of free energy in the region becomes

\begin{eqnarray}
\psi_s &=& e^{-(2D-K-J)\beta}\hspace{0.3in}\text{above } l_{11},\nonumber\\
\psi_{s}&=& 2{e^{-J\,\beta/2}}\,\,\,\,\,\,\,\,\text{below and on } l_{11}.
\label{psi1}
\end{eqnarray}
We shall label the light blue shaded sub-region of ${\bf{\{11}\}}$ between $l_3$ and $l_{11}$ as ${\bf{\{11}\}'}$.

In the region $\bf{\{31\}}$,
\begin{eqnarray}
	\psi_{s}&=& \frac{1}{2} e^{-\beta(K-2D+J)/2}.
	\label{psi3}
\end{eqnarray}

Along the line $l_1$ the singular part of free energy scales as

\begin{eqnarray}
	\psi_{s}&=& e^{-J\beta/2}  \hspace{0.75in}\text{($D > 1,K>0$) }\nonumber\\
	\psi_{s}&=& {\frac{(1+\sqrt{5})}{2}}{e^{-J\beta/2}} \hspace{0.2in}\text{($D = 1,K=0$)}\nonumber\\
	\psi_{s}&=& e^{-(8D-3J)\beta/10}  \hspace{0.45in}\text{($D < 1,K<0$)}. 
	\label{psi5} 
\end{eqnarray}

Along $l_2$ it scales as

\begin{eqnarray}
	\psi_{s}&=& {\frac{(\sqrt{5}-1)}{(5+\sqrt{5})}}{e^{-(3J-8D)\beta/9}}, 
	\label{psi6}
\end{eqnarray}

and along the line $l_3$,
\begin{eqnarray}
	\psi_{s}&=& {\frac{1}{\sqrt{5}}}{e^{-J\beta/2}}.
	\label{psi7}
\end{eqnarray}
Finally, at the triple point $\bf{T}$, the scaling of free energy goes as
\begin{eqnarray}
	\psi_{s}&=& {\frac{1}{4}}{e^{-J\beta/2}}.
	\label{psi8}
\end{eqnarray}

On the $f$- line, with reference to fig.(\ref{gs}a), the singular free energy becomes
\begin{eqnarray}
	\psi_{s}&=& {2 e^{-(J+4 K)\beta/2}}.
	\label{psi9}
\end{eqnarray}

\subsection{Scaling of the correlation lengths}
\label{fretwo}
 In a similar manner the scaling of the correlation length can also be worked out. We present some results now. On the lines $l_1$ and $l_3$ the limiting expression of correlation length $\xi_m$ becomes
\begin{equation}
\xi_m= e^{J\,\beta/2}
\label{xi1}
\end{equation}
 while on $l_2$ it becomes
\begin{equation}
\xi_m = \frac{5+\sqrt{5}}{2\sqrt{5}-2}\,e^{\beta(5J/6-8D/9)}.
\label{xi2}
\end{equation}
On the other hand the correlation length $\xi_q$ scales as the inverse of free energy everywhere on $l_1$. Its asymptotic behaviour on $l_1$ is as follows,
\begin{eqnarray}
\xi_q &=& 0 \,\,\,\,\,\,\,\,\,\,\,\,\,\,(D=3/8)\nonumber\\
\xi_q &=& \frac{1}{2}\psi_s^{-1}\,\,\,\,\,(3/8<D<1)\nonumber\\
\xi_q &=& \frac{1}{\sqrt{5}}\,\xi_m=\frac{1}{10} \left(5+\sqrt{5}\right)\psi_s^{-1}\,\,\,\,\,(D=1)\nonumber\\ 
\xi_q &= & \,\,\,\xi_m\,\,\,\,\,\, = \psi_s^{-1}\,\,\,\,\,(D>1)
\label{xi21}
\end{eqnarray}

 In the region $\bf{\{11\}}$ the limiting expression of $\xi_m$ becomes
\begin{equation}
\xi_m = \frac{1}{2} e^{\beta\,J/2}
\label{xi3}
\end{equation}
 and in $\bf{\{31\}}$ as
\begin{equation}
\xi_m\sim  e^{\beta(K-2D+2J)/2}.
\label{xi4}
\end{equation}
In the region $\bf{\{33\}}$ the scaling of correlation length varies according to a changing pattern which we could not detect fully but, nonetheless, deep enough into the $\bf{\{33\}}$ region we have checked that the limiting expression is
\begin{equation}
\xi_m =  \frac{1}{2}\, e^{9\,\beta\,J/2}.
\label{xi5}
\end{equation}
On the $f$ line the asymptotic expression for the correlation length becomes
\begin{equation}
\xi_m =  \frac{1}{2}\,e^{(J+4K)\beta/2}.
\label{xi6}
\end{equation}

Interestingly, there appear to be several instances here where the correlation length does not scale as the inverse free energy, in an apparent violation of the hyperscaling relation. 
It is noteworthy that such anomalies occur only in the $H=0$ plane. On the ``genuinely'' critical $\mathcal{L}_1$ surface hyperscaling is always followed. Significantly, as we shall see in the following, geometry efficiently encodes this anomaly via its sectional and three dimensional state space curvatures. We add that physically, this anomaly seems surprising if not intriguing since cases of hyperscaling violation normally imply that some other fluctuation is at work in addition to the usual thermal one as is the case in random field Ising models, \cite{dfisher}. Admittedly, we have not explored the apparent hyperscaling violation in any depth in this work but only made some observations of the zero field scaling of the free energy (setting $Y=0$ in eq.(\ref{sin1})) and compared it to the scaling of the correlation length(s). Our emphasis here is to investigate whether or not geometry is sensitive to such anomalies. We hope to return to this interesting issue in the future.

\section{Geometry of the spin-3/2 chain}
\label{geometry}

In this section we present our results following the geometric analysis of the spin-3/2 chain.  We discuss mainly the sectional curvatures $R_m$ and $R_q$, $R_{\omega}$ and the $3d$ scalar curvature $R_D$, in both the scaling and the non-scaling regions of the parameter space. In both cases we shall be able to amply demonstrate the power of geometry in encoding the underlying correlations in the order parameter(s). In subsections \ref{zhzw} and \ref{geozwnzh} we shall discuss, respectively, the geometry of the BEG chain and the more general $W\neq 0$ case.

\subsection{Geometry of the BEG case ($W=0$)}
\label{zhzw}


\begin{figure*}[t!]
	\begin{subfigure}[b]{0.3\textwidth}
		\centering
		\includegraphics[width=2.3in,height=1.8in]{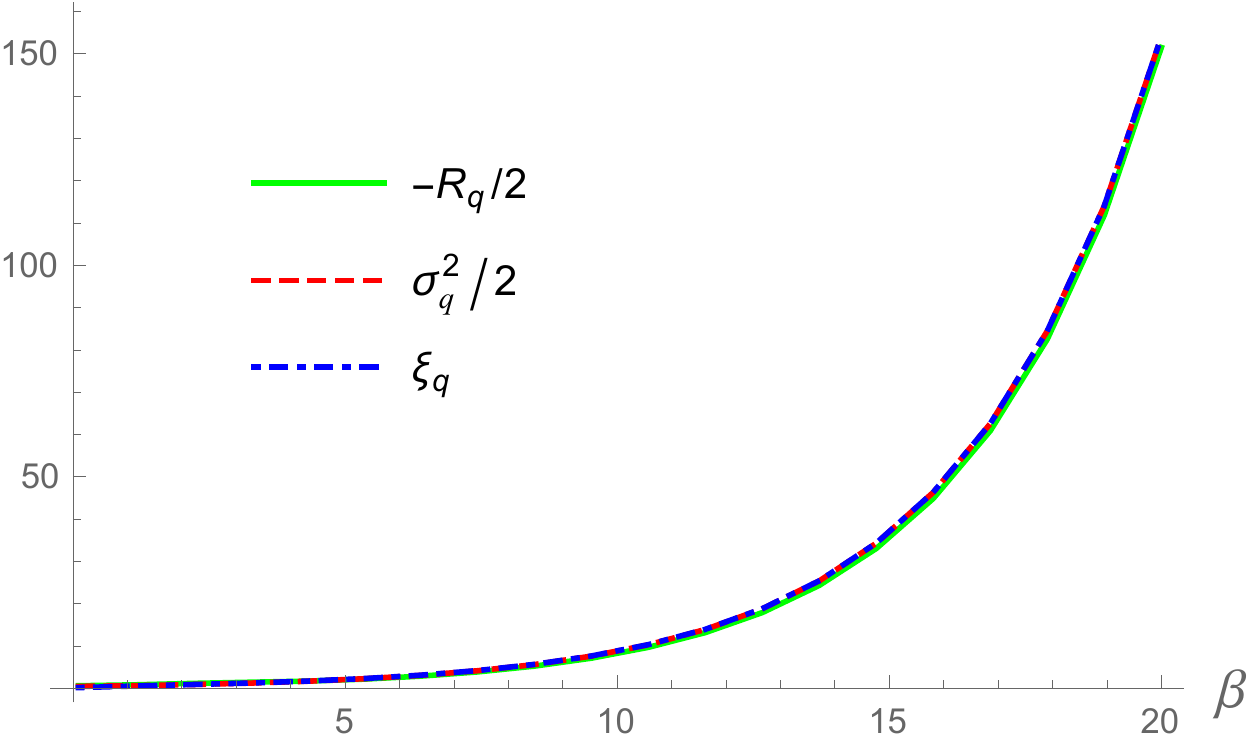}
		
		\caption{}
	\end{subfigure}
	\hspace{1.4in}
	\begin{subfigure}[b]{0.4\textwidth}
		\centering
		\includegraphics[width=2.3in,height=1.8in]{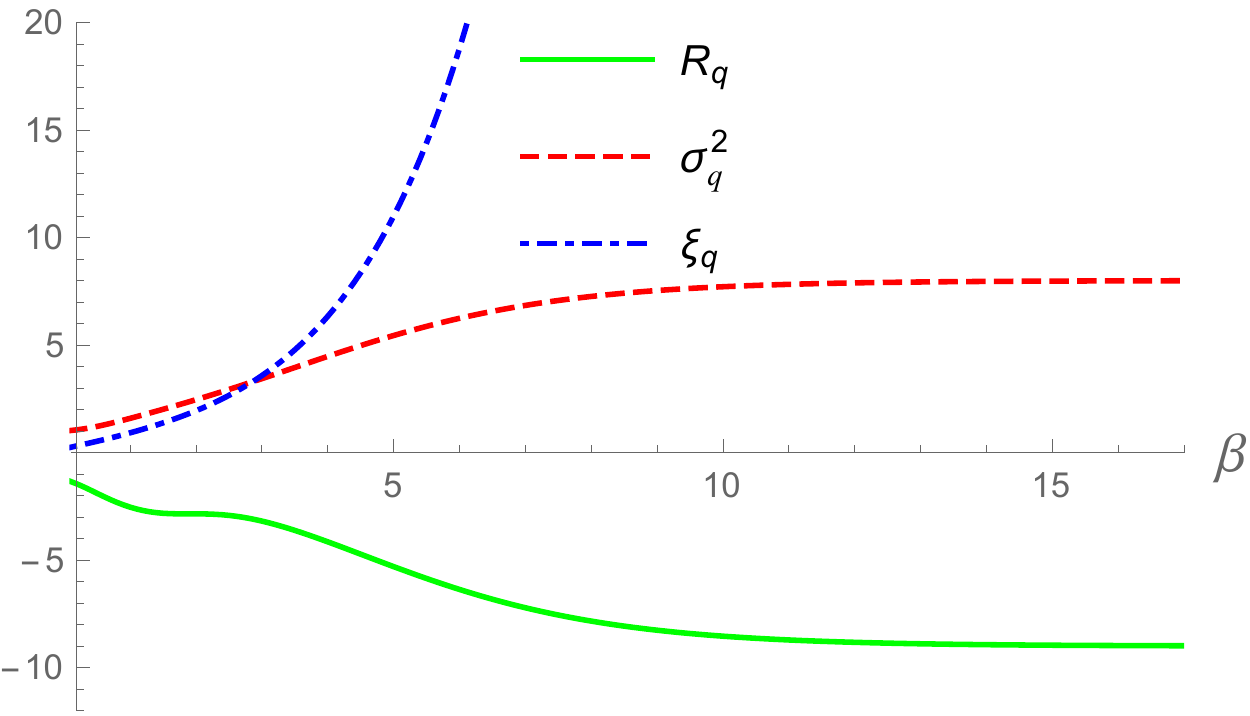}
		
		\caption{}
	\end{subfigure}
	
	\begin{subfigure}[b]{0.3\textwidth}
		\centering
		\includegraphics[width=2.3in,height=1.8in]{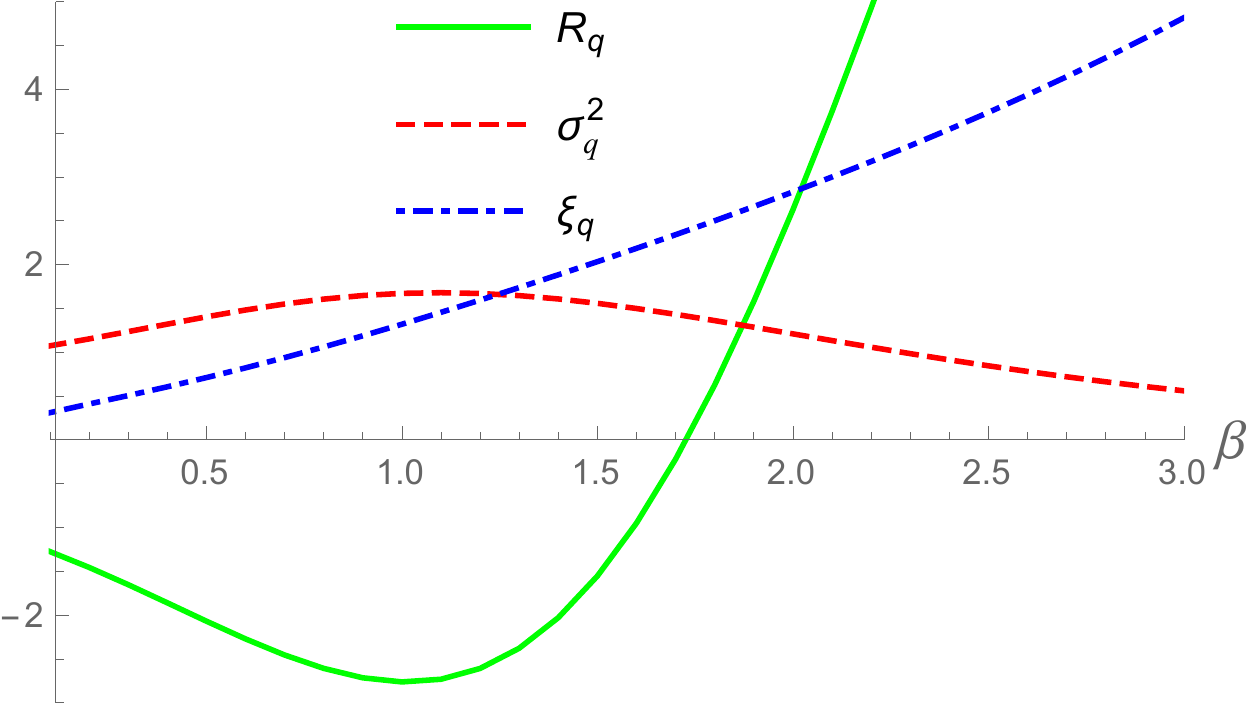}
		
		\caption{}
	\end{subfigure}
	\hspace{1.4in}
	\begin{subfigure}[b]{0.4\textwidth}
		\centering
		\includegraphics[width=2.3in,height=1.8in]{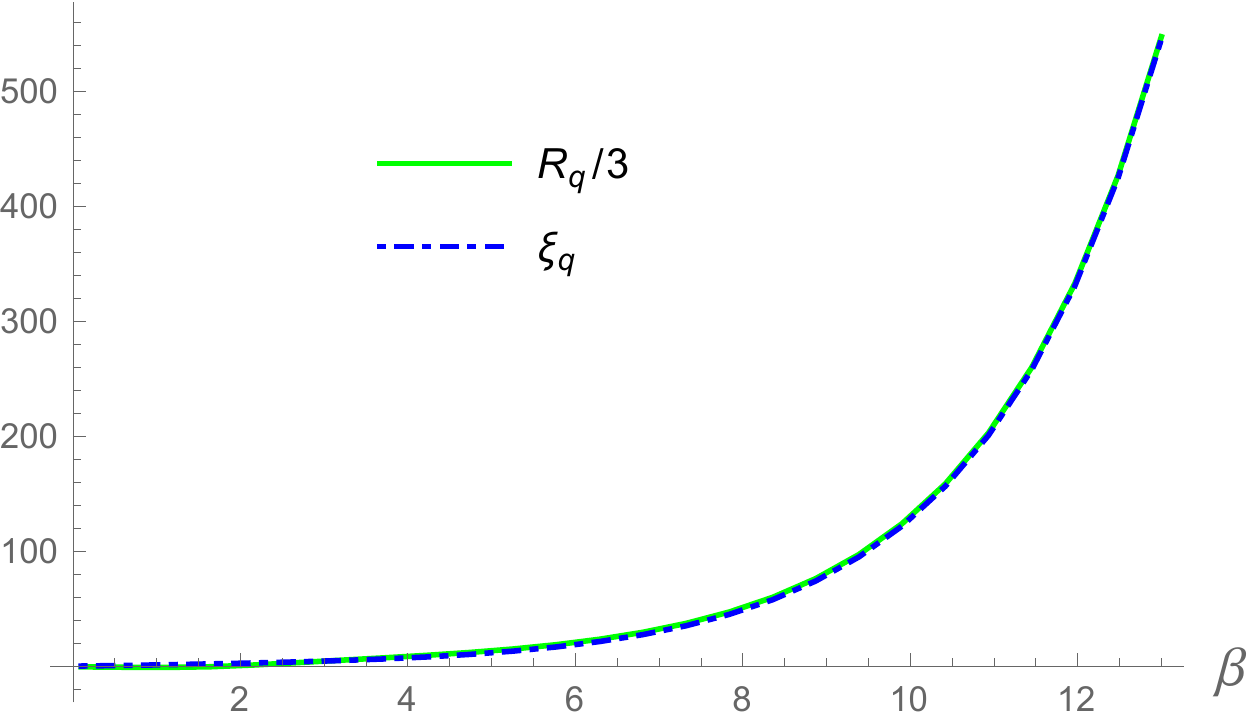}
		
		\caption{}
	\end{subfigure}
	\caption{\small{Plots with respect to $\beta$ of the sectional curvature $R_q$, mean square quadrupole fluctuation $\sigma_q^2$ and the quadrupole correlation length $\xi_q$  on the line $l_1$ for the BEG case, with $(a)$ $D=12/16,K=-1/10$,  $(b)$  $D= 21/16, K=1/8$, $(c)$  $D = 30/16,K=7/20$,  and $(d)$ same as $(c)$ but for a larger range of $\beta$.  }}
	\label{rq1}
\end{figure*}

In this subsection we present our results for the one-dimensional spin-3/2 BEG model. Following previous discussion on phase structure, we know that the spin-3/2 BEG chain already exhibits many qualitatively new features not found in the spin-one BEG chain.
 We first describe the geometry associated with the zero $H$ field, dealing separately with the coexistence lines $l_1,l_2,l_3$ and the ground state configurations $\{33\},\{13\}$ and $\{11\}$. This is followed by the geometry of the case $H\neq 0$.

\subsubsection{{\bf{Zero magnetic field, $H=0$}} }
\label{zh}

The whole of the zero field ($H=W=0$) phase diagram is a plane of symmetry for the dipole and octopole moments and is critical for their fluctuations. Correspondingly, as we will see, the sectional curvatures $R_m, R_\omega$ and the $3$-$d$ curvature $R_D$ all diverge everywhere on the symmetry plane. However, the sectional curvature $R_q$, which is expected to encode the quadrupolar correlations, does not diverge everywhere. In addition, we shall see that geometry encodes the aforementioned hyperscaling anomaly  in that $R_D$ encodes the correlation length while $R_m$ encodes the inverse free energy whenever their respective scaling behaviours do not match.

In this subsection we will focus on the detailed behaviour of the quadrupolar curvature $R_q$, the dipolar curvature $R_m$ and the $3d$ curvature $R_D$. We will first present our detailed results for the coexistence lines $l_1$, $l_2$, $l_3$ and the lines $l_{11}$ and $l_{33}$. Finally, we will outline our results for the curvatures $R_m$ and $R_D$ in parameter regions for different ground states.

\textit{{The quadrupolar curvature $R_q$}}:
We first report our results for the sectional curvature $R_q$ on the coexistence lines. On the line $l_1$, which is the only site of criticality in the quadrupolar order for the zero field BEG chain, the  asymptotic
expression of $R_q$ is as follows
\begin{eqnarray}
R_q  &=&  -1\,\,\,\,\,(D=3/8,\,\mbox{triple point})\nonumber\\
R_q  &=& \psi_s^{-1}\,\,\,\,(3/8<D\leq 1)\nonumber\\ 
R_q  &=& \frac{1}{25} (4 D+1) (16 D-21)\,\psi_s^{-1}\,\,\,(D>1)\nonumber\\ 
\label{Rql1}
\end{eqnarray}

Interestingly, $R_q$ undergoes a sign change on $l_1$ at low temperatures beyond the point ${\bf{P}}$ of fig.(\ref{phasekd1}b) with $D>21/16$ where it asymptotes to positive infinity even though it continues to scale as the inverse of free energy \footnote{Exactly at the point ${\bf{P}}$ $R_q$ asymptotes to a finite value of $-9$}. The positive divergence at criticality of the state space curvature is  in contrast to its expected negative divergence \cite{rupp1}. This anomalous behaviour of the quadrupolar curvature is related to the equally anomalous variations in quadrupolar fluctuations discussed earlier in subsection \ref{gs1}. 
We check that the mean square quadrupolar fluctuation $\sigma_q^2$ has the following asymptotic expression along $l_1$, 
\begin{eqnarray}
\sigma_q^2&=& 1\hspace{1.1in}(D=3/8)\nonumber\\
\sigma_q^2&=& e^{(8D-3)\beta/10}=\psi_s^{-1}\,\,(3/8<D\le 1)\nonumber\\
\sigma_q^2 &= & 8\, e^{(21-16D)\beta/10}\,\,\,\,\,\,\,\,\,\,\,\,\,\,(D\geq 1)
\label{qfluc}
\end{eqnarray}
Thus, for $D>1$ the divergence of quadrupole fluctuations becomes less and less steep compared to that of $\xi_q$ until beyond the point ${\bf{P}}$ at $D=21/16$ where the quadrupole fluctuations begin to decay, as mentioned earlier\footnote{At the point ${\bf{P}}$ $\sigma_q^2 \to 8$.}. Remarkably, comparing eq.(\ref{Rql1}) for $R_q$ to eq.(\ref{qfluc}) above we find that the factor $(21-16D)$ which appears in the anomalous scaling for $\sigma_q^2$ for $D>1$ also appears in the amplitude of $R_q$ for $D>1$.  We recall that beyond the point ${\bf{P}}$ the numerator in eq.(\ref{qsigma}) decays to zero faster than the numerator so that even as the correlation length $\xi_q$ continues to diverge the quadrupolar fluctuations decay. Possibly, the same underlying statistical reason that causes quadrupole fluctuations to decay also effects a sign change in $R_q$. Significantly, the scalar curvature encodes both the regular scaling of $\xi_q$ and the anomalous statistics of $\sigma_q^2$ via its magnitude and signature respectively.

 In fig.(\ref{rq1}) we show plots of $R_q$, $\sigma_q^2$ and $\xi_q$ $vs.$ $\beta$ along the line $l_1$ for a range of $D$ values. Thus, in fig.(\ref{rq1}a) where $D$ is less than one the quadupolar curvature is twice in magnitude compared to the quadrupolar correlation length. At the point ${\bf{P}}$ with $D=21/16$ in fig.(\ref{rq1}b) the quadrupolar curvature and the fluctuation moment both become constant while the correlation length continues to diverge. In fig.(\ref{rq1}c) $R_q$ changes sign to positive for $D=30/16$, but not before dipping to a negative minimum at about the same time as $\sigma_q^2$ undergoes a local maxima before the latter eventually decays to zero. As discussed earlier, the sign change in $R_q$ could possibly be related to the change in the underlying statistics of quadrupolar correlations. Fig.(\ref{rq1}d) which displays the plot in fig.(\ref{rq1}c) for a larger range of $\beta$ clearly brings out the fact that even as $R_q$ turns positive it continues to scale as the correlation length and is $153/50$ times $\xi_q$ as per eq.(\ref{Rql1}).
 
 On the lines $l_2$ and $l_3$ there is no criticality in the quadrupolar order parameter and the correlation length $\xi_q$ is seen to asymptote to the same value on both the coexistence lines,
 \begin{eqnarray}
 \xi_q &\to& \log \,\frac{1}{2} \left(3+\sqrt{5}\right)\nonumber\\
 &\sim& 0.962\,\,\,\,\,\,\,\,\,\,\hspace{0.75in}(\text{on}\, l_2\, \text{and}\, l_3)
 \label{xiql2l3}
 \end{eqnarray}
 
 Satisfyingly, dipolar curvature $R_q$ too asymptotes to small constant values on $l_2$ and $l_3$,
 \begin{eqnarray}
 R_q  &= & -\frac{1}{\sqrt{5}}\sim -0.45\hspace{0.8in}(\text{on}\,l_2)\nonumber\\
  R_q  &= & -1-\frac{1}{\sqrt{5}}\sim -1.45\hspace{0.55in}(\text{on}\,l_3)
  \label{Rqlil2}
 \end{eqnarray}
 
Note that the asymptotic values of the curvature are close to, or even smaller than, the unit lattice size which is consistent with the correlation length being of order unity.
\vspace{0.5in}

 \textit{The dipolar curvature $R_m$ and the $3d$ curvature $R_{D}$}:

It is observed that $R_m$ (and also $R_\omega$) diverges to negative infinity on all the lines $l_1$, $l_2$ and $l_3$ including the triple point and this is consistent with the divergence of $\sigma_m^2$ (and $\sigma_\omega^2$) everywhere on $H=0$ plane of the spin-3/2 BEG model. Similarly, the $3d$ curvature $R_D$, which we expect to better capture the correlations in magnetization, also diverges everywhere on the plane $H=0$. However, the scaling of the two differs in some instances on the lines and, interestingly, this variation seems to encode the hyperscaling anomaly discussed previously in section \ref{free}.

Interestingly, we observe that on $l_3$ where hyperscaling is followed both $R_m$ and $R_D$ are asymptotically equal,
\begin{eqnarray}
-R_m = -R_D = \psi_s^{-1}= 2\,\xi_m
\text{       on $l_3$}
\label{RmRDonl3}
\end{eqnarray}

 On the other hand on $l_2$ where the the spin correlation length scales faster than the inverse free energy (compare eq.(\ref{psi6}) with eq.(\ref{xi2})) the two scalar curvatures encode separate behaviours,

\begin{eqnarray}
\left.\begin{aligned}
R_m &\sim& \psi_s^{-1}\\
R_D &=& - \xi_m.
\end{aligned}\,\,\right\rbrace
\text{on $l_2$}
\label{RmRDl2}
\end{eqnarray}
Thus, while the dipolar curvature $R_m$ continues to scale as the inverse free energy the $3d$ curvature $R_D$ now encodes the spin correlation length $\xi_m$.

On the triple point we get the following asymptotic behaviour of curvatures,
\begin{equation}
R_m=R_D=\psi_s^{-1}\,\,\,\,\,\text{triple point.}
\end{equation}

Along the line $l_1$ for $3/8<D<1$ we recall that the inverse singular free energy scales as the quadrupolar correlation $\xi_q$ ( see eq.(\ref{xi21})) but not as the dipolar correlation length $\xi_m$ which diverges at a faster rate of $e^{\beta J/2}$. We could say that, as far as the quadrupolar correlations are concerned the free energy scaling is consistent with hyperscaling. For $D\geq 1$ the scaling of both $\xi_q$ and $\xi_m$ becomes the same and is consistent with the inverse singular free energy as can be checked again from eq.(\ref{xi21}). 

Along $l_1$ the asymptotic expressions/scaling behaviour of $R_m$ and $R_D$ is as follows,
\begin{eqnarray}
R_m &=& R_D =  \psi_s^{-1} \,\,\,\,\,\,\,\,\,\,\,\,\,\,\,(D=3/8)\nonumber\\
R_m &=& R_D = \kappa \psi_s^{-1} \,\,\,\,\,\,\,\,\,\,\,\,(3/8<D\leq 1)\nonumber\\
R_m &=& \frac{2}{3}\,R_D = \psi_s^{-1}\,\,\,\,\,\,\,\,\,\,(D=1)\nonumber\\
R_m & = & \kappa_1 R_D = \kappa_2 \psi_s^{-1}\,\,\,\,(D>1),
\end{eqnarray}
where, in the last case for $D>1$ the proportionality constants $\kappa_1$ and $\kappa_2$ are always of order unity. For the case $D<1$ the constant $\kappa$ is $D$ dependent and is anomalously high for $D$ near the triple point. Thus, near $D=7/16$ it is about $100$, but reduces quickly to order unity values for $D>8/16$. We see that on $l_1$ where hyperscaling is broadly followed (albeit with different correlation lengths) both $R_m$ and $R_D$ have a similar scaling behaviour as the inverse of free energy (though sometimes with an anomalously large proportionality constant).

\begin{figure}[!h]
	\centering
	\includegraphics[width=2.5in,height=2.2in]{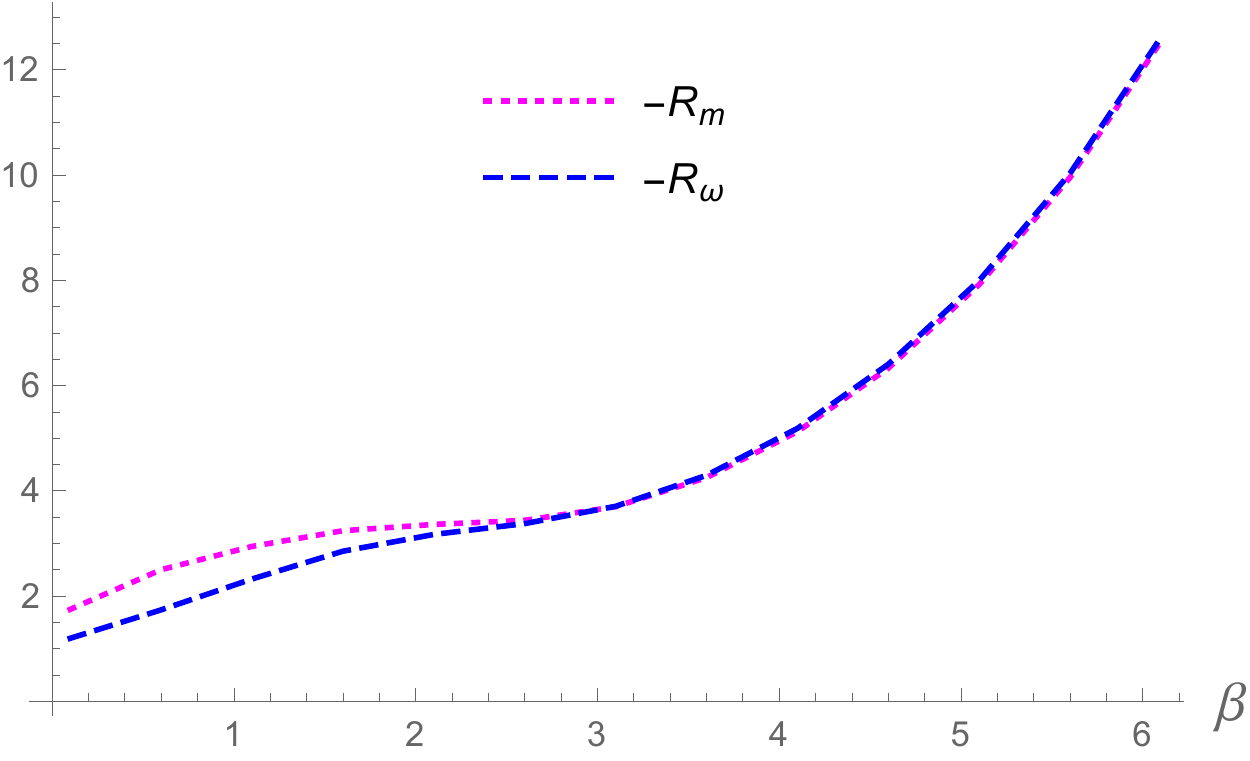}
	\caption{\small{ Plot of $R_m$, $R_{\omega}$ on the line $l_1$ with $H=0$, $W=0$, $K= -1/40$ and $ D=15/16$}}
	\label{rmrozhzw}
\end{figure}

In fig.(\ref{rmrozhzw}) we plot the dipolar curvature $R_m$ and the octopolar curvature $R_{\omega}$ on $l_1$ to demonstrate that the two sectional curvatures approach each other. Given that the correlation lengths of the dipole and octopole moments are the same and their statistics are similar the equality of their respective sectional curvatures further substantiates our approach of representing correlations via suitable hypersurface geometries.

\begin{figure*}[t!]
	\begin{subfigure}[b]{0.3\textwidth}
		\centering
		\includegraphics[width=2.3in,height=1.8in]{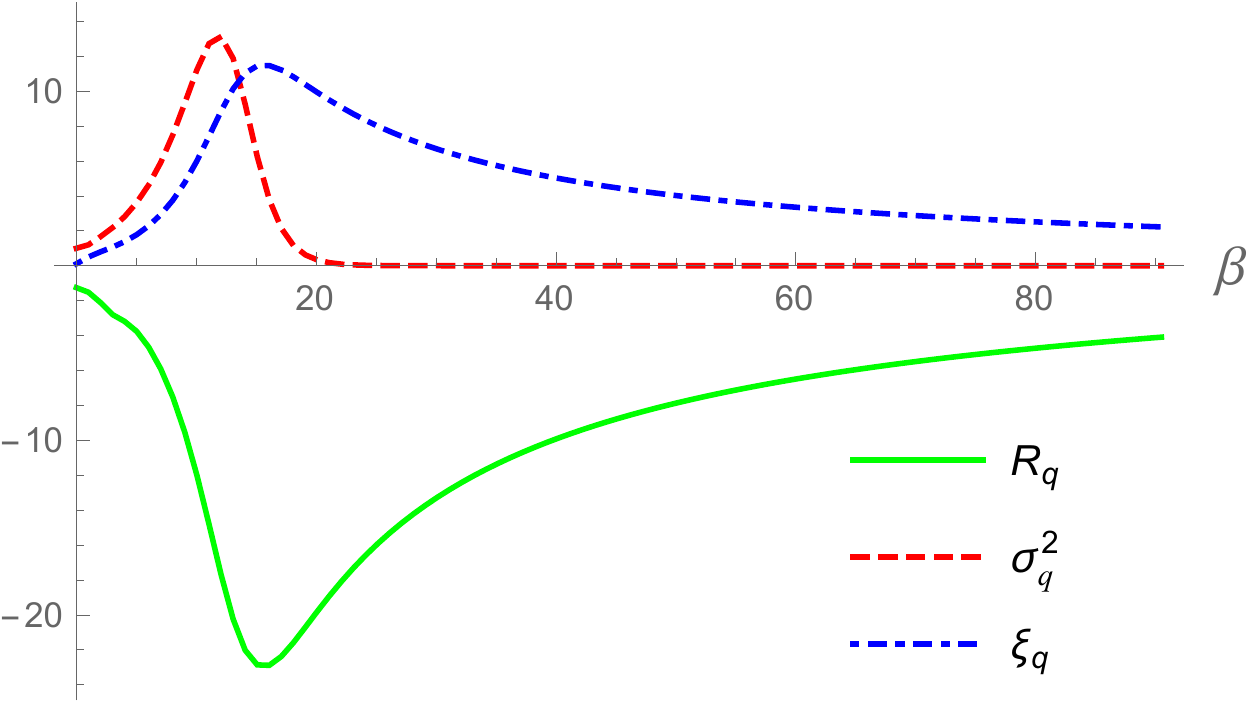}
		
		\caption{}
	\end{subfigure}
	\hspace{1.4in}
	\begin{subfigure}[b]{0.4\textwidth}
		\centering
		\includegraphics[width=2.3in,height=1.8in]{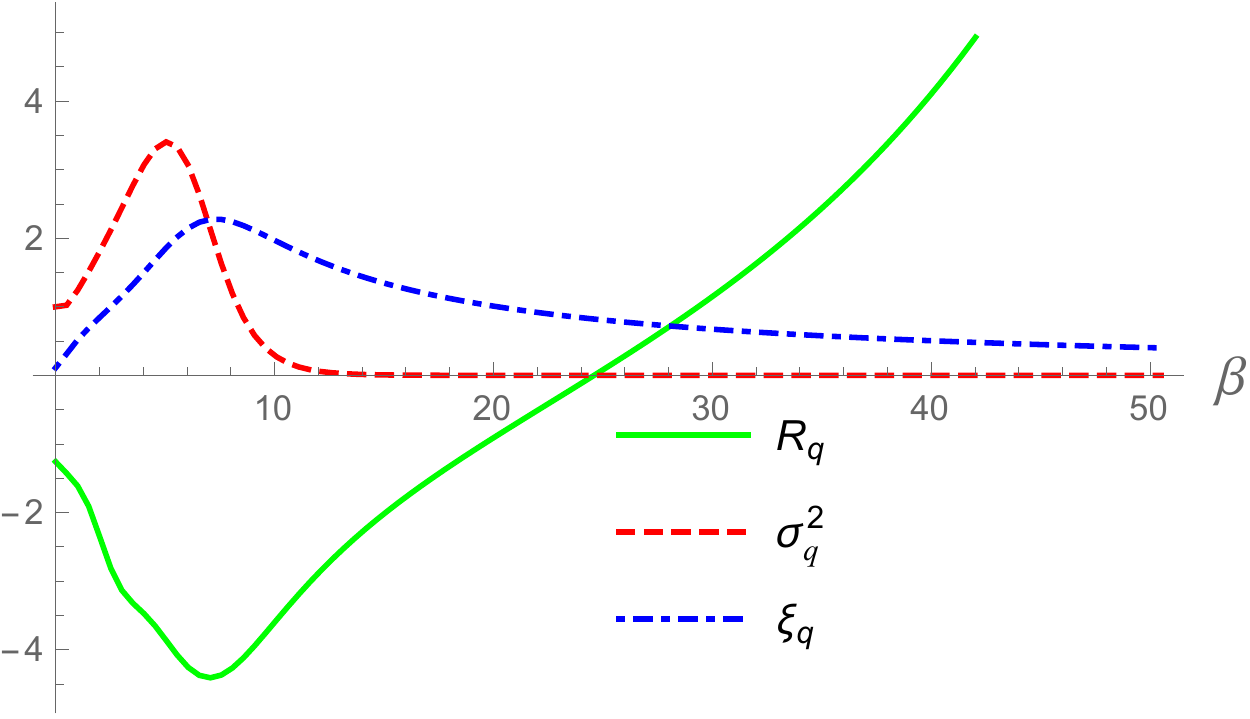}
		
		\caption{}
	\end{subfigure}
	\caption{\small{Plots of $R_q$, $\xi_q$ and $\sigma_q^2$ for spin-3/2 BEG chain at $H=W=0$ in the {\bf{\{33\}}} region above the $l_1$ line. . In $(a)$ the plot is slightly above the $l_1$ line with $K=-1/8+1/1000$ and $D=11/16$ and in $(b)$ it is slightly above $(a)$ with  $K=-1/8+1/100$ and $D=11/16$. }}
	\label{Rq33}
\end{figure*}

\begin{figure*}[t!]
	\begin{subfigure}[b]{0.3\textwidth}
		\centering
		\includegraphics[width=2.3in,height=1.8in]{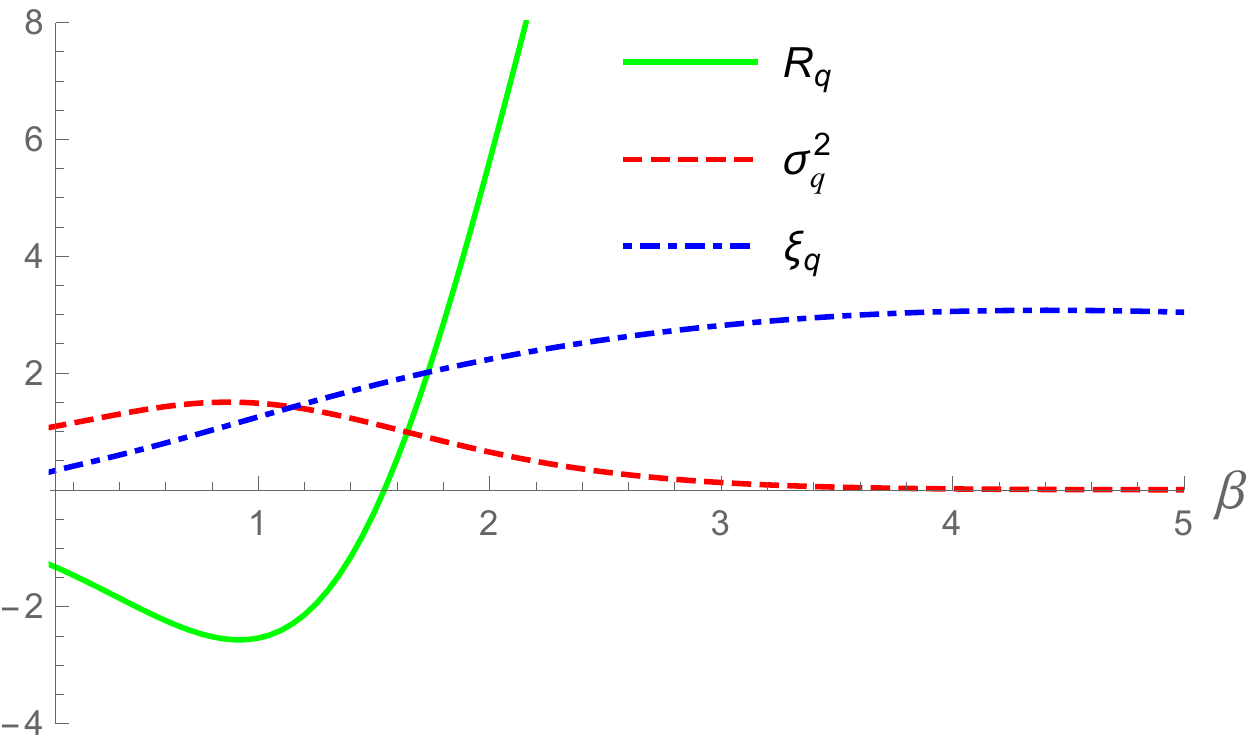}
		
		\caption{}
	\end{subfigure}
	\hspace{1.4in}
	\begin{subfigure}[b]{0.4\textwidth}
		\centering
		\includegraphics[width=2.3in,height=1.8in]{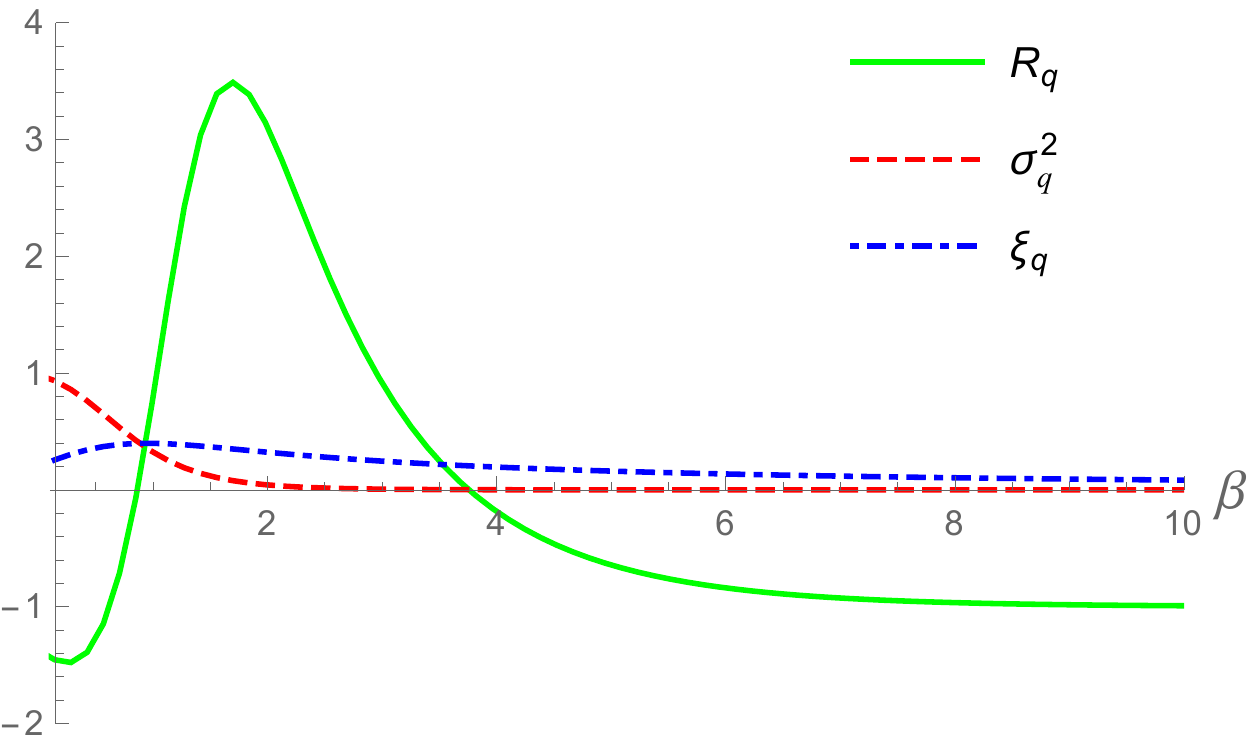}
		
		\caption{}
	\end{subfigure}

	\caption{\small{Plots of $R_q$, $\xi_q$ and $\sigma_q^2$ for spin-3/2 BEG chain at $H=W=0$ in the {\bf{\{11\}}} region below the $l_1$ line. . In $(a)$ the plot is slightly below the $l_1$ line with $K=7/20-1/100$ and $D=30/16$ and in $(b)$ it is below $(a)$ with  $K=7/20-1/3$ and $D=30/16$. }}
	\label{Rq11}
\end{figure*}

\begin{figure*}[t!]
	\begin{subfigure}[b]{0.3\textwidth}
		\centering
		\includegraphics[width=2.3in,height=1.8in]{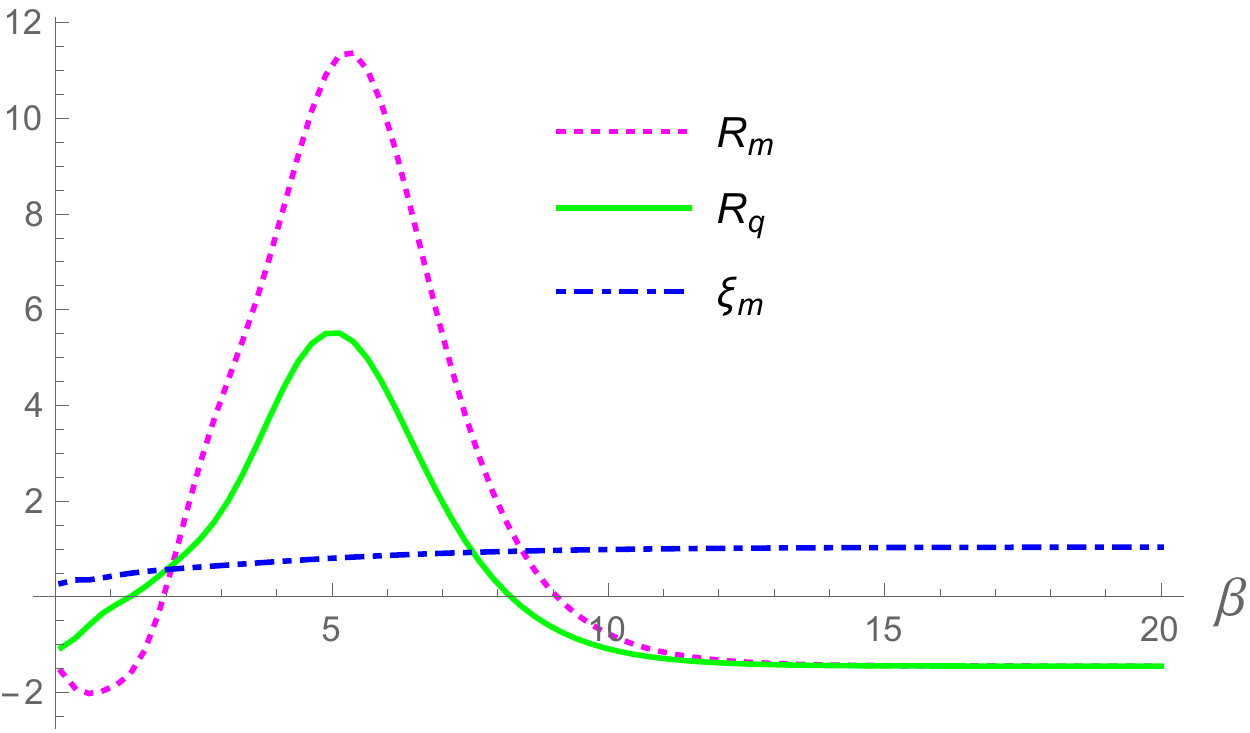}
		
		\caption{}
	\end{subfigure}
	\hspace{1.4in}
	\begin{subfigure}[b]{0.4\textwidth}
		\centering
		\includegraphics[width=2.3in,height=1.8in]{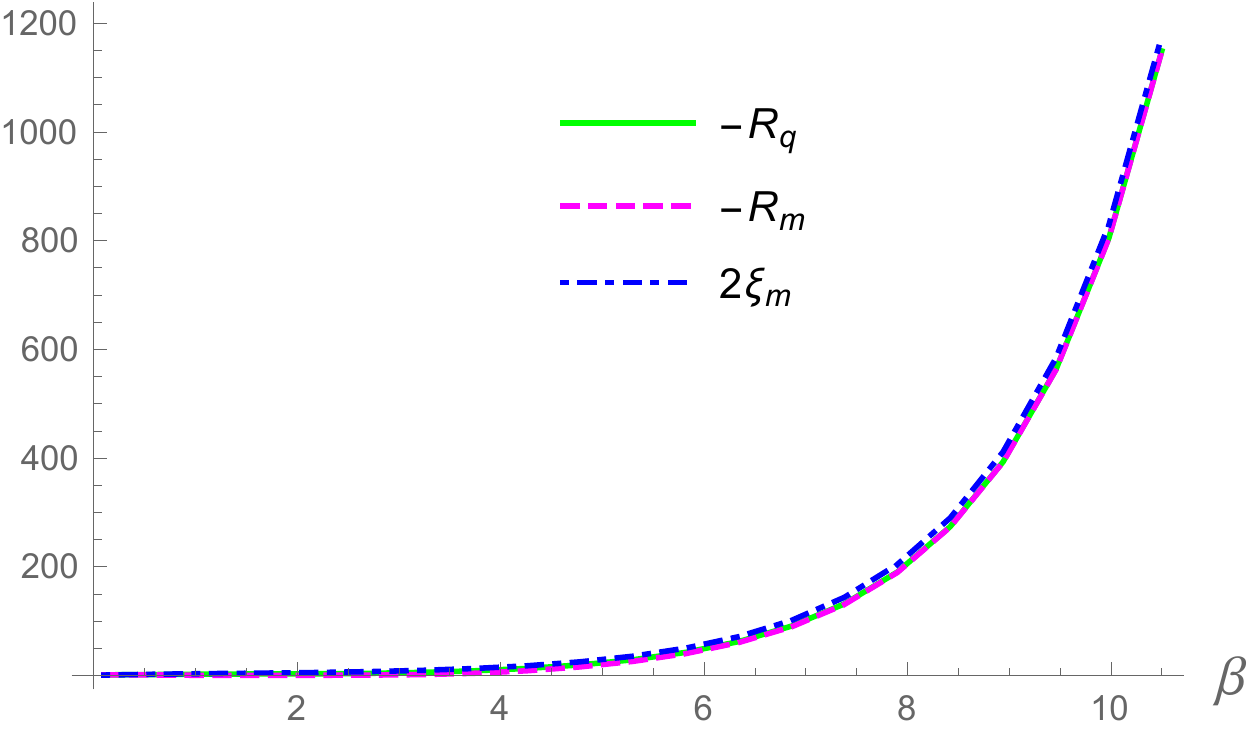}
		
		\caption{}
	\end{subfigure}
	\caption{\small{ $(a)$. Plots of $R_m$, $R_q$ and $\xi_m$ on the coexistence line $l_2$ for $W=0$ and $H=0.1$ at $D=23/60,K-1/3.$   $(b)$. Plot of $-R_m,-R_q$ and $\xi_m$ on the $l_1$ line for $H=0.1$ and $D=13/10,K=1/10$. }}
	\label{nzH1}
\end{figure*}

\textit{ $R_m$, $R_q$ and $R_D$ in the regions}:
Recalling that the plane $H=0$ remains critical for spin (and octopolar) fluctuations, we now report our observations for $R_m$ and $R_D$ in the parameter regions corresponding to the stable ground state configurations. Quite satisfyingly, we find that in all regions where hyperscaling is followed, namely ${\{\bf{33}\}}$ above and on $l_{33}$ and ${\{\bf{11}\}}$ below and on $l_{11}$ in fig.(\ref{psiphase}),  the curvatures $R_m$ and $R_D$ both scale as the inverse of free energy. On the other hand, where it is not followed, namely in ${\{\bf{13}\}}$, ${\{\bf{33}\}'}$ and ${\{\bf{11}\}'}$, the curvature $R_m$ continues to scale as the inverse free energy while $R_D$ goes as the correlation length $\xi_m$. To summarize, therefore, we have
\begin{equation}
R_m \sim  R_D\sim \psi_s^{-1}\sim \xi_m,\text{ \small{ where hyperscaling holds}}
\label{RmRDhyper}
\end{equation}
and
\begin{eqnarray}
\left.\begin{aligned}
R_m &\sim& \psi_s^{-1}\\
R_D &\sim&  \xi_m.
\end{aligned}\,\,\right\rbrace
\text{\small{where hyperscaling fails.}}
\label{RmRnonhyper}
\end{eqnarray}

Moving forward to our observation of the quadrupolar curvature in the regions we show in fig.(\ref{Rq33}a) plots of $R_q$, $\xi_q$ and $\sigma_q^2$ in the region ${\bf{\{33\}}}$ slightly above the $l_1$ line and then follow it up with fig.(\ref{Rq33}b) for a similar plot a bit higher above the $l_1$ line. We see that close to the line $R_q$ approaches small negative values, almost mirroring the correlation length $\xi_q$  while sufficiently deeper into the region it flips to positive values in the ${\bf{\{33\}}}$ region and undergoes a positive divergence. Possibly, the dipolar curvature is again encoding a change in the statistics of fluctuations here.  In fig.(\ref{Rq11}a) and fig.(\ref{Rq11}b) respectively we plot $R_q$, $\xi_q$ and $\sigma_q^2$ in the region ${\bf{\{11\}}}$ slightly below the $l_1$ line and then go further deep into the region. This time the geometry seems to suggest the opposite behaviour. Thus, close to the $l_1$ line in the ${\bf{\{11\}}}$ region $R_q$ changes sign to positive and diverges indicating a change in statistics of quadrupole correlations. On the other hand, deeper into the region it switches to negative and asymptotes to $-1$ thus suggesting that `normal' correlation statistics prevails at low enough temperatures.

\subsubsection{{\bf{Non-zero magnetic field, $ H\neq 0$}} }
\label{nzhzw}

\begin{figure*}[t!]
\begin{subfigure}[b]{0.3\textwidth}
\centering
	\includegraphics[width=2.3in,height=1.8in]{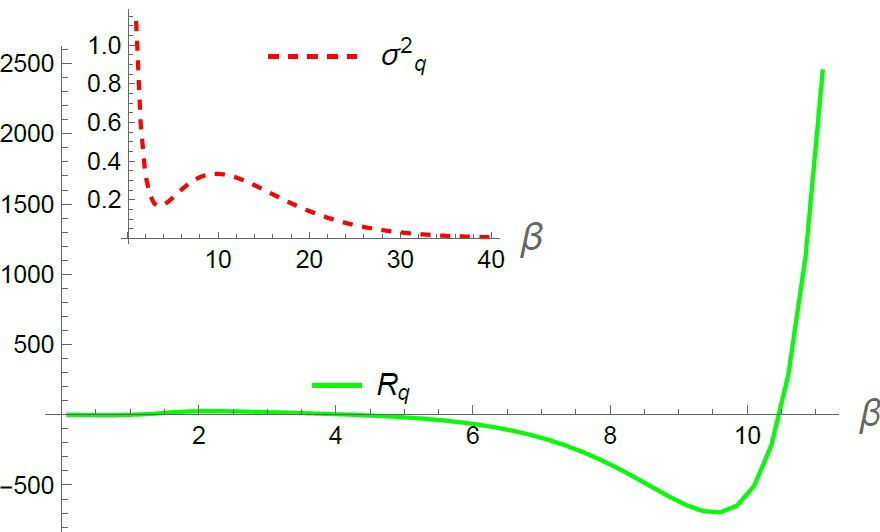}
	
		\caption{}
			\end{subfigure}
		\hspace{1.4in}
		\begin{subfigure}[b]{0.3\textwidth}
			\centering
			\includegraphics[width=2.3in,height=1.8in]{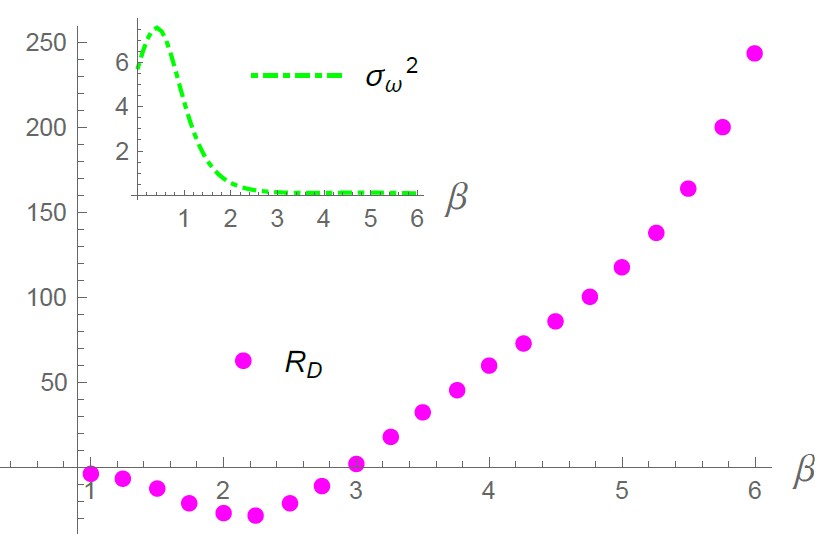}
			
			\caption{}
	\end{subfigure}

\caption{\small{ $(a)$ Plot with of  $R_q$ with respect to $\beta$, and  $\sigma_q^2$ vs $\beta$ in the inset, for $H=0.1,K=0.62,D=2.6$ and $W=0$. $R_q$ has a minima at approximately the same tempertaure as the $\sigma_q^2$ maxima and it flips to positive as the latter begins to decay. $(b)$ Plot with of  $R_D$ with respect to $\beta$, and  $\sigma_\omega^2$ vs $\beta$ in the inset, for $H=0.02,K=0.62,D=2.56$ and $W=0$. $R_D$  flips to positive as $\sigma_\omega^2$ begins to decay.  }}
	\label{nzH2a}
\end{figure*}

We recall from discussions around fig.(\ref{gs}b) that for non-zero $H$ the $l_2$ and $l_3$ coexistence lines are non-critical while $l_1$ is critical upto an $H$ dependent ${\bf{P}}$ point beyond which the spin and quadrupole fluctuations decay even as the correlation length continues to diverge (see eq.(\ref{mlinemoments}) and the $m$ lines in fig.(\ref{khatualines}a)). In fig.(\ref{nzH1}a) we show a plot of $R_m$ $R_q$ and $\xi_m$ on the $l_2$ coexistence line for $H=0.1$. It is seen that both the sectional curvatures closely follow each other for low temperatures and asymptote to a value of $-1.5$ while the correlation length limits to $1$. Exactly the same behaviour is seen on the coexistence line $l_1$, not depicted here. In fig.(\ref{nzH1}b) we show a plot of $R_m,R_q$ and $ \xi_m$ somewhere on the $l_1$ line at a point before the ${\bf{P}}$ point and for $H=0.1$. They are clearly seen to merge with each other with curvatures double the correlation length asymptotically. At other points and for other values of $H$ the proportionality constant $R_m/\xi_m$ might change but is always close to $1$.

We now move to a geometric description of the ${\mathcal{L}_1}$ surface near the $m$ lines of fig.(\ref{khatualines}a) for values of the quadrupolar coupling $K$  greater than $0.515$ (see eq.(\ref{kcritical})). As mentioned previously, the order parameter fluctuations decay below the $m$ lines while $\xi_m$ diverges everywhere. We carefully observe the sectional curvature $R_q$ along with $\sigma_q^2$ for a value of $H$ below the $m_2$ line for $K>0.515$.  In fig.(\ref{nzH2a}a) we plot $R_q$ vs. $\beta$ and $\sigma_q^2$ vs. $\beta$ in the inset at $H=0.1$ on the $K=0.62$ line of fig.(\ref{khatualines}a). We find that instead of a negative divergence proportional to $\xi_m$ as is the case above the $m$ lines, the curvature $R_q$ now changes sign and turns positive at about the same $\beta$ as $\sigma_q^2$ drops down after a local peak. In other words, the ability of $R_q$ to encode anomalous decay of fluctuations beyond the ${\bf{P}}$ point on the $l_1$ line in the zero field case (see fig.(\ref{rq1}c) and discussion around eq.(\ref{qsigma})) continues for the respective ${\bf{P}}$ points  in the non-zero field case. Notice once again, as in fig.(\ref{rq1}c) for $H=0$ the negative peak of $R_q$ is at the same place ($\beta=10$) as the positive peak of $\sigma_q^2$. For both zero and non-zero field cases we suspect that the statistics of correlations changes at around the temperature where $R_q$ turns positive. Also, note that unlike the zero field case the statistics of dipole and quadrupole fluctuations are similar in the non-zero field case. 

We add that the sectional curvature $R_m$ is not sensitive to change in the statistics of fluctuations across the $m$ lines and continues its negative divergence proportional to $\xi_m$ both above and below the $m$ lines. Significantly the $3d$ curvature $R_D$ which is expected to encode dipolar and octopolar correlations, does respond to a change in fluctuation statistics below the $m$ lines. In fig.(\ref{nzH2a}b) $R_D$ is seen to change sign to positive at about the same place as the octopolar fluctuation moment (shown in the inset) begins to decay. Once again, the statistics of dipole, quadrupole and octopole fluctuations are similar in the non-zero field case. 

\begin{figure*}[t!]
\begin{subfigure}[b]{0.3\textwidth}
\centering
	\includegraphics[width=2.3in,height=1.8in]{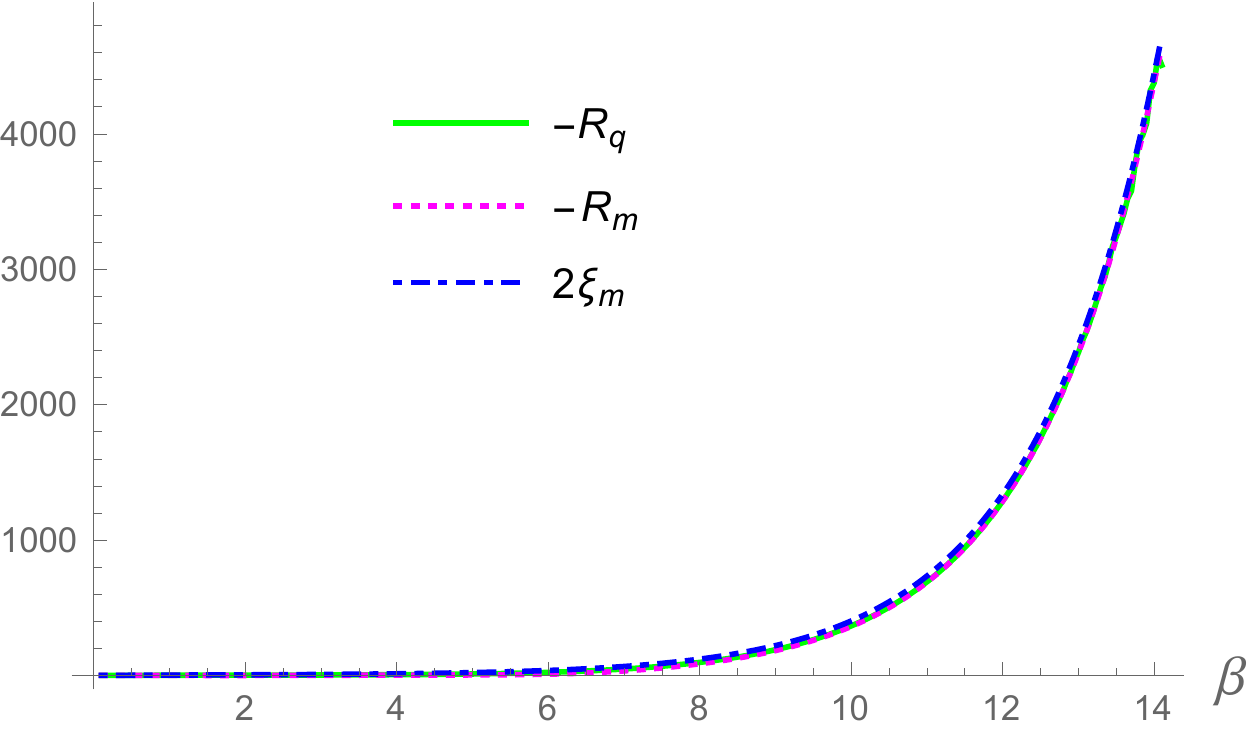}
	
		\caption{}
			\end{subfigure}
		\hspace{1.4in} 
		\begin{subfigure}[b]{0.3\textwidth}
			\centering
			\includegraphics[width=2.3in,height=1.8in]{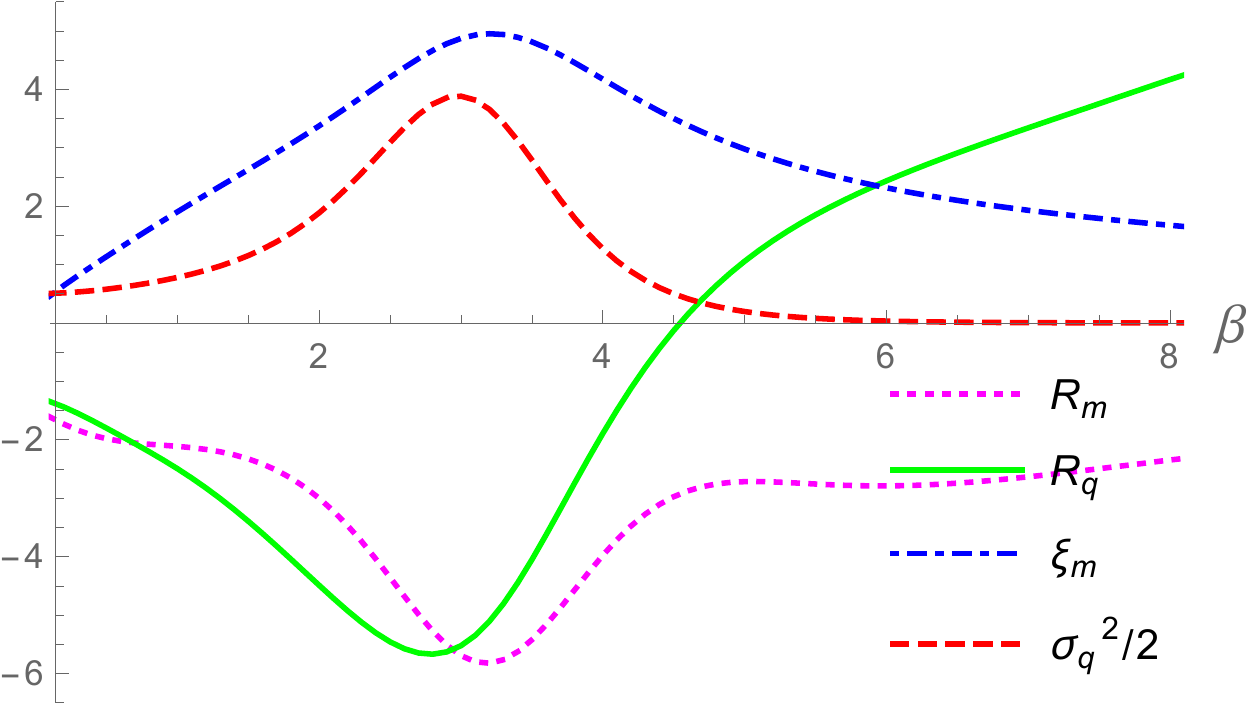}
			
			\caption{}
	\end{subfigure}

\caption{\small{ Plot of $R_m$,  $R_q$ ,  $\xi_m$ for $H=0.01$, $W=0.02$, $K=0.05 $ with $(a)$  $D = 1.1625$ on $l_1$ line  and $(b)$ $D=1.125$ in the ${\bf{\{33\}}}$ region.  }}
\label{nzw1}
\end{figure*}

\begin{figure*}[t!]
\begin{subfigure}[b]{0.3\textwidth}
\centering
	\includegraphics[width=2.3in,height=1.8in]{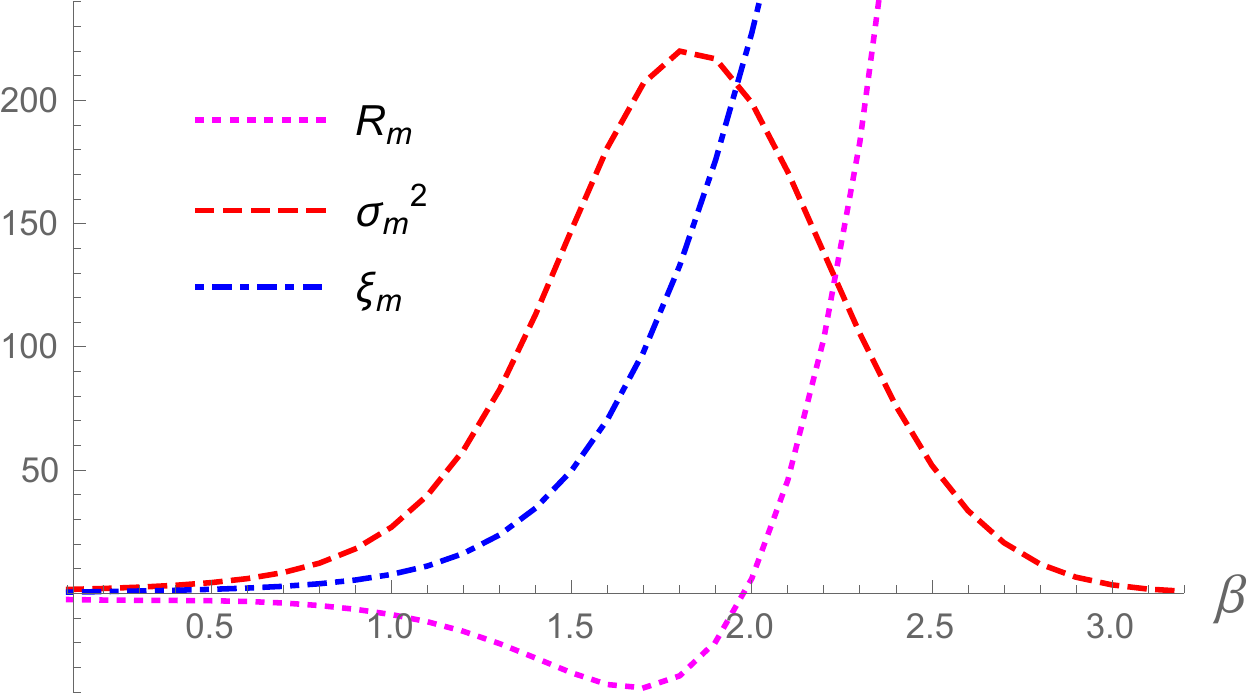}
	
		\caption{}
			\end{subfigure}
		\hspace{1.4in}
		\begin{subfigure}[b]{0.3\textwidth}
			\centering
			\includegraphics[width=2.3in,height=1.8in]{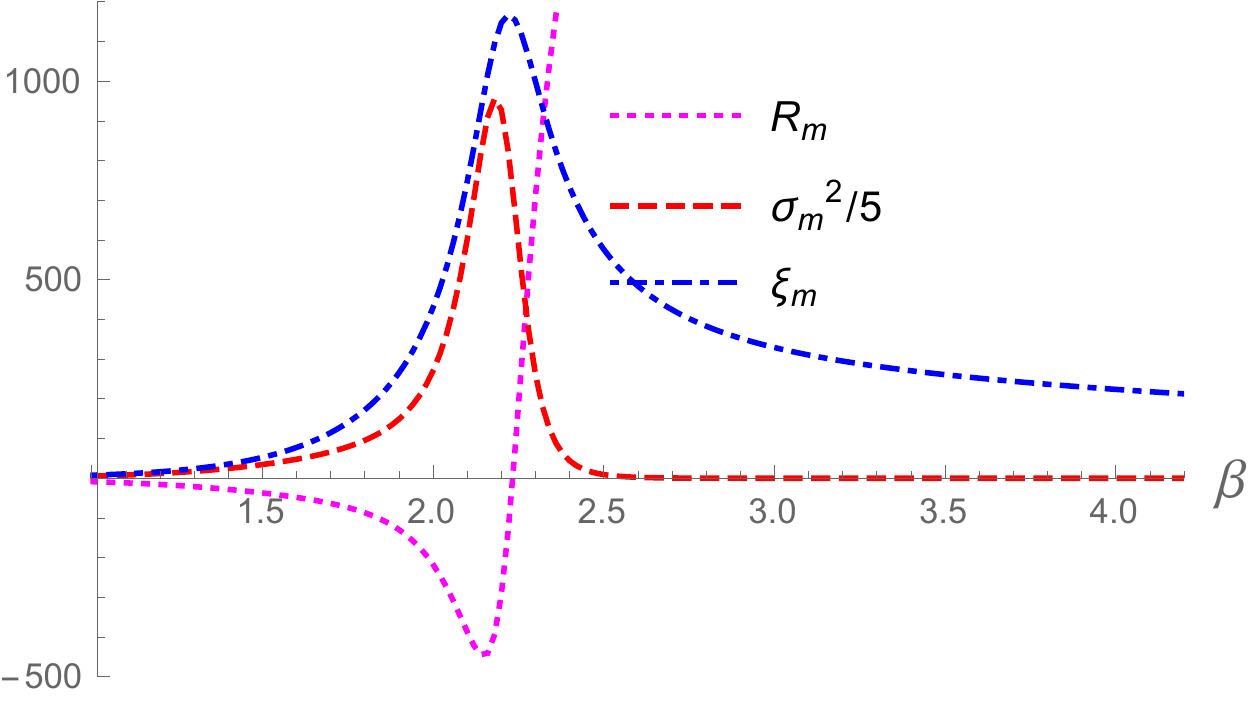}
			
			\caption{}
	\end{subfigure}

\caption{\small{$(a)$ Plot of $R_m$, $\xi_m$ and $\sigma_m^2$ for $H=0.3$, $K=1$, $D=3$, $W=-0.1333$. Compare with fig.(\ref{negative W}a) which plots $m$ vs. $\beta$ for the same parameters. $(b)$ Plot of $R_m$, $\sigma_m^2$ and  $\xi_m$ for $H=0.3$, $W=-0.1335$, $K=1/8$ and $D=6/8$. Compare with fig.(\ref{negative W}b) which plots $m$ vs. $\beta$ for the same parameters.}}
\label{nzw2}
\end{figure*}

\subsection{ Geometry of the general case ($W\neq0$) }
\label{geozwnzh}

For the general spin-3/2 chain with $W\neq 0$ we have seen earlier that the phase structure remains mostly qualitatively similar as is borne out by a comparison fig.(\ref{khatualines}b) for $W=1/10$ with fig.(\ref{khatualines}a) for $W=0$. Without going into the details of scaling, etc we report that the geometry of the non-BEG spin chain is qualitatively similar to the BEG case, with the behaviour of $R_m$, $R_q$ and $R_D$ along similar lines in similar parameter regions. In fig.(\ref{nzw1}a) we plot the sectional curvatures and the correlation length on the $l_1$ line where we see that both $R_m$ and $R_q$ coincide and are in a fixed ratio to the correlation length, with the proportionality constant of order unity.

 In fig.(\ref{nzw1}b) we show plots of $R_m, \xi_m,R_q$ and $\sigma_q^2$ in the  ${\bf{\{33\}}}$ region. $R_m$ is seen to mirror the correlation length $\xi_m$. On the other hand, $R_q$ flips to positive at around the same place as the quadrupole fluctuation substantially decays. Once, again it seems here that $R_q$ seems to be encoding a change in quadrupole correlation statistics. 

Finally, we report briefly on the geometry of the  interesting case of opposite signs of $H$ and $W$ as discussed in eq.(\ref{W negative}) and represented through magnetization plots in fig.(\ref{negative W}a) and fig.(\ref{negative W}b). In fig.(\ref{nzw2}a) we plot the dipolar sectional curvature $R_m$ and the correlation length $\xi_m$ along with the spin fluctuation moment $\sigma_m^2$ for parameter values same as for the magnetization plot in  fig.(\ref{negative W}a). Here, the magnitude of negative $W$ is a little les  than its `critical' value for the given $H$ (see eq.(\ref{kcritical})) so that the magnetization still remains positive though it makes a rapid transition at a finite $\beta$ as is evident from fig.(\ref{negative W}a). Here we see a case of anomalous decay of spin fluctuations even as the correlation length continues to diverge. $R_m$ this time is indeed sensitive to the change and accordingly it flips sign to positive at about the same place as $\sigma_m^2$ begins its decay. This is reminiscent of similar behaviour of $R_q$ in several places.

In fig.(\ref{nzw2}b) we plot  $R_m$ and $\xi_m$ for the same parameter values as fig.(\ref{negative W}b) where the magnetization flips sign to $-3/2$ at lower temperatures owing to a stronger energy-lowering influence of $W$ as compared to $H$. As is clear from fig.(\ref{nzw2}) there is an accompanying sharp rise in the correlation length $\xi_m$ followed by a more relaxed decay to zero at lower temperatures. Interestingly, the curvature $R_m$ too undergoes a negative peak of about the same magnitude\footnote{The ratio between the $R_m$ peak and the $\xi_m$ peak remains order one in all cases.} but then it flips to positive values and diverges. Thus, this becomes an instance of the dipolar curvature $R_m$ suggesting a change of statistics in a manner different from erstwhile cases. Here it seems to be taking a cure from the changing behaviour of the correlation length itself.

\section{Conclusions}
\label{conclu32}

In this work we have investigated in detail the phase structure and the state space geometry of the one-dimensional spin-3/2 lattice model both for the case when the Hamiltonian is BEG and for a more general case with the octopolar coupling field $W$ turned on. On several occasions geometry has been able to shine a light on the interesting phase behaviour of the spin-3/2 chain.

 This work finds interesting aspects of the ground state phase behaviour of the spin-3/2 chain not reported elsewhere. Thus, for both the BEG  and the general case, the critical $\mathcal{L}_1$ surface extending in the $H-D-K$ parameter space in fig.(\ref{phasekd1}a) has portions where the order parameter fluctuations anomalously decay while the correlation length continues to diverge as represented in fig.(\ref{khatualines}).  Furthermore, we have also documented region wise variations in the scaling of singular free energy as represented in fig.(\ref{psiphase}). Yet another interesting, if curious, observation is that in several regions of the parameter space the hyperscaling relation is apparently violated, in that the inverse of the singular free energy scales slower than the correlation length.
 
We investigate in detail the dipolar sectional curvature $R_m$, the quadrupolar sectional curvature $R_q$, and the $3d$ curvature $R_D$. Our geometrical investigations amply confirm  Ruppeiner's strong as well as weak conjecture. Namely, near the (pseudo)critical point the scalar curvature is found to be equal to the correlation length upto order unity constant and away from criticality the appropriate scalar curvature corresponds well with a decaying or asymptotically small correlation length. Satisfyingly, the sectional curvatures chosen on appropriate surfaces are seen to efficiently encode correlations in the corresponding order parameters. The sectional curvature method, employed earlier in the context of Kerr-AdS black holes in \cite{asknbads} and for the spin-one chain in \cite{riekan1}, thus appears to be a robust geometrical means to probe systems with multiple order parameters.

Significantly, we have been able to systematically document the sign change in sectional curvatures with the onset of anomalous fluctuation behaviour mentioned above. Thus, at about the same temperatures as the fluctuations in an order parameter begin to decay (despite a continued divergence in the correlation length) the relevant sectional curvature flips sign from negative to positive. In the zero-field BEG case we could also succeed in obtaining an asymptotic expression for the amplitude of $R_q$ which changes sign exactly as the scaling of the quadrupole fluctuation moment becomes negative. All this ties well with the previous, long-standing assertions about the signature of the scalar curvature encoding a change in statistics form `statistically attractive' (as in the case of critical point, fluids, Bose gases) to statistically repulsive (as in solids like states, Fermi gases, etc), see eg. \cite{ruppajp}. The anomalous decay of fluctuations, mathematically represented by a faster decay of the numerator compared to the denominator in eq.(\ref{qfluc}) or eq.(\ref{qsigma}) of course also has statistical undertones, since the aforementioned equation are themselves a result of statistical averaging over correlations of all length scales. Further, geometry seems to encode the aforementioned hyperscalinlg anomaly between the singular free energy and the correlation length. For the zero-field BEG case we have reported that while the sectional curvature $R_m$ continues to scale as the inverse of free energy, the $3d$ curvature $R_D$ encodes the correlation length. On the other hand where hyperscaling is followed the two curvatures become asymptotically equal or proportional. It will indeed be worthwhile investigating these interesting issues in the future.

 In this work we have not presented results from our ongoing work on the geometry of the mean-field spin-3/2 lattice model. We shall soon report on some interesting results of the mean-field case.
 
 \begin{acknowledgements}
 
 We gratefully acknowledge useful discussions with Tapobrata Sarkar, Suresh Govindarajan and George Ruppeiner.
 
 \end{acknowledgements}

\end{document}